\documentclass[]{entalpic}

\usepackage{amsmath}
\usepackage{amssymb}
\usepackage{graphicx}
\usepackage[numbers]{natbib}
\usepackage[most]{tcolorbox}

\usepackage{enumitem}
\setlist{topsep=5pt, partopsep=5pt, itemsep=5pt, parsep=5pt}

\usepackage[utf8]{inputenc} % allow utf-8 input
\usepackage[T1]{fontenc}    % use 8-bit T1 fonts
\usepackage{hyperref}       % hyperlinks
\usepackage{url}            % simple URL typesetting
\usepackage{booktabs}       % professional-quality tables
\usepackage{amsfonts}       % blackboard math symbols
\usepackage{nicefrac}       % compact symbols for 1/2, etc.
\usepackage{microtype}      % microtypography
\usepackage{xcolor}         % colors
\usepackage{todonotes}
\usepackage{multirow}
\usepackage{amsmath}
\usepackage{siunitx}
\usepackage{caption}
\usepackage{graphicx}
\usepackage{bm}% bold math
\usepackage{subcaption}
\usepackage{listings}
\usepackage[version=4]{mhchem}
\usepackage{booktabs}
\usepackage{amssymb}      % for \checkmark
\usepackage{longtable}
\usepackage{amsmath}
\usepackage{array}      % for p{} column types
\usepackage[table]{xcolor} % for \rowcolor in tables
\usepackage{chngcntr}
\usepackage{array}
\usepackage{multirow}

\makeatletter
\renewcommand{\title}[1]{\gdef\@title{#1}\newcommand{\titlelist}{%
  {\LARGE\bfseries\rmfamily\raggedright\linespread{1.1}\selectfont #1\par}%
  \vspace{0.4cm}%
}}
\makeatother

\title{LeMat-Bulk: aggregating, and de-duplicating quantum chemistry materials databases}

\author[1,*]{Martin Siron}
\author[1]{Inel Djafar}
\author[1]{Ali Ramlaoui}
\author[1]{Etienne du Fayette}
\author[1]{Amandine Rossello}
\author[2,3]{\\Edvin Fako}
\author[4]{Matthew McDermott}
\author[5]{Felix Therrien}
\author[6]{Luis Barroso-Luque}
\author[7]{Flaviu Cipcigan}
\author[7]{\\Philippe Schwaller}
\author[8]{Thomas Wolf}
\author[1,*]{Alexandre Duval}

\affiliation[1]{Entalpic, Paris, France}
\affiliation[2]{LIAC, EPFL, Lausanne, Switzerland}
\affiliation[3]{NCCR, EPFL, Lausanne, Switzerland}
\affiliation[4]{Newfound Materials, Houston, TX, USA}
\affiliation[5]{MILA, Quebec, Canada}
\affiliation[6]{FAIR at Meta, Menlo Park, CA, USA}
\affiliation[7]{IBM, Daresbury, UK}
\affiliation[8]{Huggingface, Brooklyn, NY, USA}
\contribution[*]{Corresponding authors}

\abstract{The rapid expansion of materials science databases has driven machine learning-based discovery while also posing challenges in data integration, duplication, and interoperability. Robust standardization and de-duplication methods are needed to address these issues and streamline materials research. We present \textit{LeMat-Bulk}, a unified dataset combining Materials Project, OQMD, and Alexandria, encompassing over 5.3 million PBE-calculated materials and also representing the largest collection of PBESol and SCAN functional calculations. Our methodology standardizes calculations across databases that utilize different parameters, effectively addressing redundancy and enhancing cross-compatibility. To de-duplicate, we propose a hashing function which we termed the Bonding Algorithm Weisfeiller-Lehman (BAWL). We comprehensively benchmark this fingerprint under atomic noise, lattice strain, and symmetry transformations, demonstrating that it outperforms existing fingerprinting techniques such as SLICES, and CLOUD in robustness while offering greater computational efficiency than similarity-based approaches such as Pymatgen’s StructureMatcher. Additionally, the fingerprint facilitates the analysis of functional-dependent trends (PBE, PBESol, SCAN) offering a scalable framework for data-driven materials science.}

\begin{document}

\maketitle

\setlength{\parindent}{20pt}

\section{\label{sec:level1}Introduction}
Recent advancements in generative materials science aim to accelerate the discovery of novel materials with target properties by leveraging deep learning models trained on atomic-scale data~\citep{jiao2023crystal, park2024has, ai4science2023crystal, zeni2023mattergen,merchant2023scaling}. These AI-driven approaches offer the potential to reshape the design process by generating entirely new material structures with tailored functionalities~\citep{decost2020scientific}. In contrast to traditional heuristics and rules, generative models can explore vast design spaces and reduce human bias, providing a promising path for data-driven materials discovery and inverse design. However, the performance and reliability of these models depend heavily on access to high-quality and well-structured datasets, as well as on robust methods for evaluating materials novelty -- both of which remain open challenges.

Over the past decades, computational chemistry and in-silico modeling have played an essential role in advancing materials science.  Density Functional Theory (DFT) \citep{kohn1996density} has enabled the replication and rationalization of known material behaviors, marking the transition to the third paradigm of materials science~\citep{schleder2019dft,hafner2006toward}. This was followed by the emergence of large-scale, high-throughput computational databases such as the Materials Project~\citep{jain2013commentary, jain2020materials}, OQMD~\citep{saal2013materials}, and Alexandria~\citep{schmidt2021alexandria}. These initiatives have shifted the field toward data-centric workflows -- an evolution sometimes described as the fourth paradigm of materials science~\citep{himanen2019dataparadigm} -- where data generation, curation, and reuse are foundational to discovery. While Machine Learning (ML) plays an increasingly important role in this paradigm, it builds on the infrastructure of large, structured datasets derived from ab initio simulations. However, even with these resources, major limitations persist due to inconsistent and inhomogeneous data.

A central issue is the lack of interoperability between major quantum chemistry datasets, which prevents efficient integration and utilization. Each repository varies in calculation parameters, computational features, and data scope, which complicates cross-referencing and integration. These discrepancies can lead to inconsistencies in key thermodynamic and structural properties, which complicates the construction of ML-ready datasets and impairs the ability to draw global insights~\citep{hegde2020reproducibility}. 
Moreover, duplicate entries — arising from multiple calculations of the same structure — are often not linked or identified, further reducing the efficiency of data usage. For ML researchers, this lack of harmonization presents a barrier to benchmarking and evaluating whether generated structures are truly novel or simply rediscovered. It also creates hurdles to access and unify data across multiple sources, often requiring expertise in multiple database interfaces. Standardization efforts like the Optimade protocol~\citep{evans2024developments, andersen2021optimade} have attempted to address this issue by enabling federated querying across databases. Yet, such efforts remain limited in both scope and adoption. For instance, the Materials Project’s Optimade endpoint currently focuses on a curated subset of properties~\citep{MatSciOptimade}, while broader coverage is typically delivered through the dedicated Materials Project API. Additionally, endpoint providers often label similar properties with different names. This highlights limitations in the federated approach compared to a fully unified dataset strategy, as pursued in this study.

Efforts to reconcile discrepancies in calculation methodologies led to the Alexandria database, integrating structures from sources such as Materials Project (MP) and AFLOW~\citep{curtarolo2012aflow} to enhance interoperability and standardize parameters. Similarly, OQMD incorporates Materials Project data~\citep{griesemer2025wide}. However, these efforts lack transparency regarding data provenance (e.g., in Alexandria, material origins—MP, AFLOW, or prototype generation—are unclear), limiting broader applicability. For example, we cannot compare an Alexandria material taken from MP with its MP-computed counterpart. Besides, these initiatives often duplicate rather than consolidate computations. As a result, despite increasing data availability, standardized, interoperable, and openly accessible databases remain scarce.

Beyond data integration, a key yet unresolved challenge in generative materials discovery is defining material novelty in a computationally rigorous and scalable way. Current high-throughput pipelines frequently generate duplicate structures, yet there are no universally accepted benchmarks for determining whether generated materials are genuinely novel or already exist in databases. Existing methods for structural comparison, including widely used approaches such as Pymatgen's Structure Matcher~\citep{ong2015materials}, as well as similarity metrics proposed, such as Widdowson et al.'s Average Minimum Distance~\citep{widdowson2022average,widdowson2022resolving}, and Yang et al.\citep{yang2014proposed}, Bartók et al.\citep{bartok2013representing}, De et al.\citep{de2016comparing}, Zhu et al.\citep{zhu2016fingerprint}, Oganov et al.\citep{oganov2009quantify}, and Therrien et al.\citep{therrien2020matching}. However these metrics require pairwise structure comparisons across large datasets, resulting in computational inefficiencies at scale. A more scalable alternative is to use hashing or fingerprinting methods to represent structures as compact strings, enabling efficient direct comparisons rather than computing vector-based similarities with thresholds. Recent hashing approaches include SLICES~\citep{xiao2023invertible}, CLOUD~\citep{xu2024cloud}, and a graph-based hashing method proposed by Ongari et al.~\citep{ongari2022data}. These methods are each discussed in detail in the Supplementary Information (SI). While Ongari's method showed promise for identifying similar Metal-Organic Framework (MOF) structures, it has not yet been systematically evaluated on large inorganic crystalline materials databases, nor have CLOUD and SLICES. Moreover, there is currently no comprehensive benchmarking assessing the robustness or sensitivity of these hashing and similarity scoring methods under structural perturbations or symmetry transformations, nor have these methods been directly compared.

This methodological gap has led to growing criticisms within the generative materials domain, particularly regarding the treatment of disordered materials~\citep{merchant2023scaling, leeman2024challenges, cheetham2024artificial}. While materials science has traditionally defined materials based on their chemical composition, atomic structure, properties and applications the emergence of generative models has blurred these boundaries, further complicating how novelty should be framed and evaluated.

To address these issues, we introduce \textit{LeMat-Bulk}, a large-scale unified dataset that consolidates data from the Materials Project, OQMD, and Alexandria by standardizing field formats, curating metadata and resolving cross-compatibility issues. In addition to creating a comprehensive materials repository, we propose a revision of an existing fingerprinting method by Ongari et al.~\citep{ongari2022data}, previously used for Metal-Organic Frameworks (MOFs). We dub this modified fingerprint the Bonding Algorithm Weisfeiller-Lehman (BAWL) and use it to systematically identify duplicates and structurally similar materials. We benchmark BAWL against existing structural comparison methods: SLICES~\citep{xiao2023invertible}, 
% and a modified version of SLICES dubbed 
CLOUD~\citep{xu2024cloud}, similarity methods including Pymatgen's and Mattergen's StructureMatcher \citep{zeni2023mattergen, ong2013python}, and GNN embeddings from EquiformerV2 \citep{liao2023equiformerv2} and PDD~\citep{widdowson2022average, widdowson2022resolving}. To assess their robustness and novelty resolution, we test these methods under various perturbations, such as atomic position noise, lattice parameter variations, translations, and symmetry operations; as well as on a curated benchmark of disordered structures.

By offering both a unified dataset and validated comparison tools, LeMat-Bulk lays the foundation for scalable, reproducible, and interpretable generative workflows in materials science. Beyond accelerating data-driven discovery, this framework can help translate computationally generated candidates into experimentally actionable insights. For example, by identifying genuinely novel structures for synthesis, or validating model predictions against known benchmarks. As the field advances toward closed-loop design and autonomous experimentation, tools for dataset integration, deduplication, and novelty assessment will be critical. LeMat-Bulk aims to be one such enabling resource for the community.

\section{Methods}

\subsection{Combining datasets in \textit{LeMat-Bulk}}
We ensured cross-compatibility among three widely used databases by adopting consistent pseudopotentials, functionals, Hubbard U parameters, and spin-polarization settings. The integration of Alexandria and Materials Project was relatively seamless, as Alexandria’s computational parameters were largely modeled after those of the Materials Project~\citep{schmidt2021crystal}. We excluded records from the Materials Project due to pseudopotential discrepancies: Alexandria employs VASP's V\_sv pseudopotential for vanadium, whereas the Materials Project uses VASP's V\_pv. Additionally, Alexandria utilizes updated Cs pseudopotentials that differs from those of the Materials Project. The final discrepancy was for ytterbium Yb, whereas the Materials Project transitioned from VASP's Yb\_2 to Yb\_3, but no \ce{Yb}-containing materials were present in the dataset at the time of integration. Alexandria employed VASP's \ce{Yb}. To access Alexandria records we utilized their Optimade endpoint, while for Materials Project we utilized the Materials Project API, filtering out deprecated calculations.

The integration of OQMD posed greater challenges due to notable differences in computational parameters~\citep{shen2022reflections, schmidt2022dataset,saal2013materials}. For Hubbard U corrections, discrepancies were observed in all cases except cobalt (\ce{Co}), necessitating the exclusion of many of their records. Specifically, materials containing fluorine (\ce{F}) in combination with \ce{Co}, \ce{Cr}, \ce{Fe}, \ce{Mn}, \ce{Mo}, \ce{Ni}, \ce{V}, or \ce{W}, and materials containing oxygen (\ce{O}) in combination with \ce{V}, \ce{Cr}, \ce{Mn}, \ce{Fe}, \ce{Ni}, \ce{Cu}, \ce{Th}, \ce{U}, \ce{Np}, \ce{Pu}, \ce{Mo}, or \ce{W}, were excluded. Regarding spin-polarization calculations, OQMD did not apply spin polarization to materials without d- or f-electrons. Consequently, materials calculated without spin polarization, composed solely of s- or p-electrons were also excluded to ensure compatibility with Alexandria and Materials Project. For pseudopotentials we dropped materials containing the following elements due to incompatibility: \ce{Ca}, \ce{Ti}, \ce{V}, \ce{Cr}, \ce{Mn}, \ce{Ru}, \ce{Rh}, \ce{Ce}, \ce{Eu}, \ce{Yb}. An overview of chosen pseudopotentials and Hubbard U parameters in \textit{LeMat-Bulk} can be found in the SI (Table \ref{tab:pseudopotentials} and Table \ref{tab:hubbardu}). OQMD and Materials Project both had multiple calculations for each structure, for both we utilized the final relaxation. More details can be found in SI. Despite differences in k-point sampling and plane-wave cutoff energy settings, no materials were excluded on this basis, as all databases were assumed to operate within convergence. To access OQMD records we downloaded the SQL dump labeled v1.6 from their website~\citep{OQMDDownload}. Finally, we split these properties among PBE~\citep{perdew1996generalized}, PBESol~\citep{perdew2007generalized}, and SCAN~\citep{sun2015strongly} functionals when available.

To standardize the representation of structural data, we adopted the Optimade API specification~\citep{andersen2021optimade, evans2024developments}. This framework offers a standardized approach for defining structural parameters within quantum chemistry databases. However, Optimade serves as a federated model and does not resolve inconsistencies in property definitions across databases. For instance, Alexandria’s magnetic moment property is denoted as `\texttt{alexandria\_magnetic\_moment}', whereas other databases might use terms such as `\texttt{magmoms}.' To address such disparities, we adopted standardized naming conventions for common material properties, facilitating uniformity and interoperability within the integrated dataset (a list of all property names and data types can be found in Table \ref{tab:lemat}). Moreover, each database host determines the scope and nature of properties exposed via their Optimade endpoint.

\newcommand{\groupheading}[1]{%
  \specialrule{\lightrulewidth}{0pt}{0pt} 
  \rowcolor{gray!20}%
  \multicolumn{4}{l}{\textbf{#1}}\\[-0.1ex] 
  \specialrule{\lightrulewidth}{0pt}{0pt} 
}

\begin{longtable}{%
  >{\raggedright\arraybackslash}p{3.3cm} 
  >{\raggedright\arraybackslash}p{3.5cm} 
  >{\raggedright\arraybackslash}p{2cm} 
  >{\raggedright\arraybackslash}p{6.5cm} 
}
\caption{LeMat-Bulk Dataset Card. The names of the feature of the dataset, following by their data type (e.g. List) and shape if applicable (e.g. (3x3), whether the fields are Optimade fields as listed in their API specs, and a short description of what the field means.\label{tab:lemat}}\\

\toprule
\textbf{Feature} & \textbf{Data Type} & \textbf{Optimade field} & \textbf{Description}\\
\midrule
\endfirsthead

\multicolumn{4}{r}{\small \textit{(Table \ref{tab:lemat} continued)}}\\[4pt]
\toprule
\textbf{Feature} & \textbf{Data Type} & \textbf{Optimade field} & \textbf{Description}\\
\midrule
\endhead

\bottomrule
\multicolumn{4}{r}{\small \textit{(Continued on next page)}}\\
\endfoot

\bottomrule
\endlastfoot

\groupheading{Cell components}

lattice\_vectors 
  & (List[List[Float]], 3x3) 
  & \checkmark 
  & Lattice matrix (e.g.\ [[3,0,0],[0,3,0],[0,0,3]]) \\

cartesian\_site\_positions 
  & (List[List[Float]], Nx3) 
  & \checkmark 
  & Cartesian coordinates \\

species\_at\_sites 
  & (List[String]) 
  & \checkmark 
  & List of elements at each site (e.g.\ ['Li','O','O']) \\

material\_fingerprint 
  & (String) 
  & 
  & Fingerprint based on BAWL algorithm \\

\groupheading{Composition}

chemical\_formula\_anonymous
  & (String) 
  & \checkmark 
  & Anonymous formula (e.g.\ A$_2$B for O$_2$Li$_4$) \\

chemical\_formula\_descriptive
  & (String) 
  & \checkmark 
  & Descriptive formula (e.g.\ Na(H$_2$O)$_6$) \\

chemical\_formula\_reduced
  & (String)
  & \checkmark
  & Reduced formula (e.g.\ Li$_2$O for O$_2$Li$_4$) \\

elements
  & (List[String])
  & \checkmark
  & List of elements (e.g.\ [``Li'',``O'']) \\

species
  & (JSON)
  & \checkmark
  & Optimade species format (concentration, mass, etc.) \\

elements\_ratios
  & (Dictionary)
  & \checkmark
  & Fractional element ratios (e.g.\ \{``Li'':0.33, ``O'':0.67\}) \\

\groupheading{Other Optimade queryable properties}

nelements
  & (Integer)
  & \checkmark
  & Distinct element count (e.g.\ 2 for Li$_4$O$_2$) \\

nsites
  & (Integer)
  & \checkmark
  & Total sites (e.g.\ 6 for Li$_4$O$_2$) \\

nperiodic\_dimensions
  & (Integer)
  & \checkmark
  & Number of repeated dimensions (e.g.\ 3) \\

dimension\_types
  & (List[Integer], 3x1)
  & \checkmark
  & Periodic dimensions (e.g.\ [1,1,1]) \\

immutable\_id
  & (String)
  & \checkmark
  & Unique ID (e.g.\ oqmd-YYY) \\

last\_modified
  & (Date/time)
  & \checkmark
  & Modification date \\

\groupheading{Structure-wide properties}

energy
  & (Float)
  & 
  & Total system energy (eV) \\

total\_magnetization
  & (Float)
  & 
  & Total system magnetization ($\mu_B$) \\

stress\_tensor
  & (List[List[Float]], 3x3)
  & 
  & 3x3 stress tensor (kB) (e.g.\ [[3,0,0],[0,1,0],[0,0,1]]) \\

\groupheading{Site-specific properties}

forces
  & (List[List[Float]], 3xN)
  & 
  & Force per site (eV/\AA) \\

magnetic\_moments
  & (List[Float])
  & 
  & Magnetic moment ($\mu_B$) \\

\groupheading{Other metadata}

cross\_compatibility
  & (Boolean)
  & 
  & DFT compatibility flag (True if entries are cross compatible) \\

functional
  & (String: pbe, pbesol, scan)
  & 
  & DFT functional (pbe, pbesol, scan) \\
\end{longtable}

For the purpose of stability assessment, and to compute above the hull energies, we corrected PBE energies using the recently proposed  `110 PBE' method~\citep{rohr2024simple}. This energy correction scheme aims to mitigate systematic errors in PBE-calculated total energies. Hull energies were computed using Pymatgen’s phase diagram module, allowing us to assess the thermodynamic stability of materials relative to competing phases.

\subsection{BAWL: a graph-based fingerprint method for inorganic crystal structures}

\begin{figure}[htbp]
    \centering
    \includegraphics[width=0.8\textwidth]{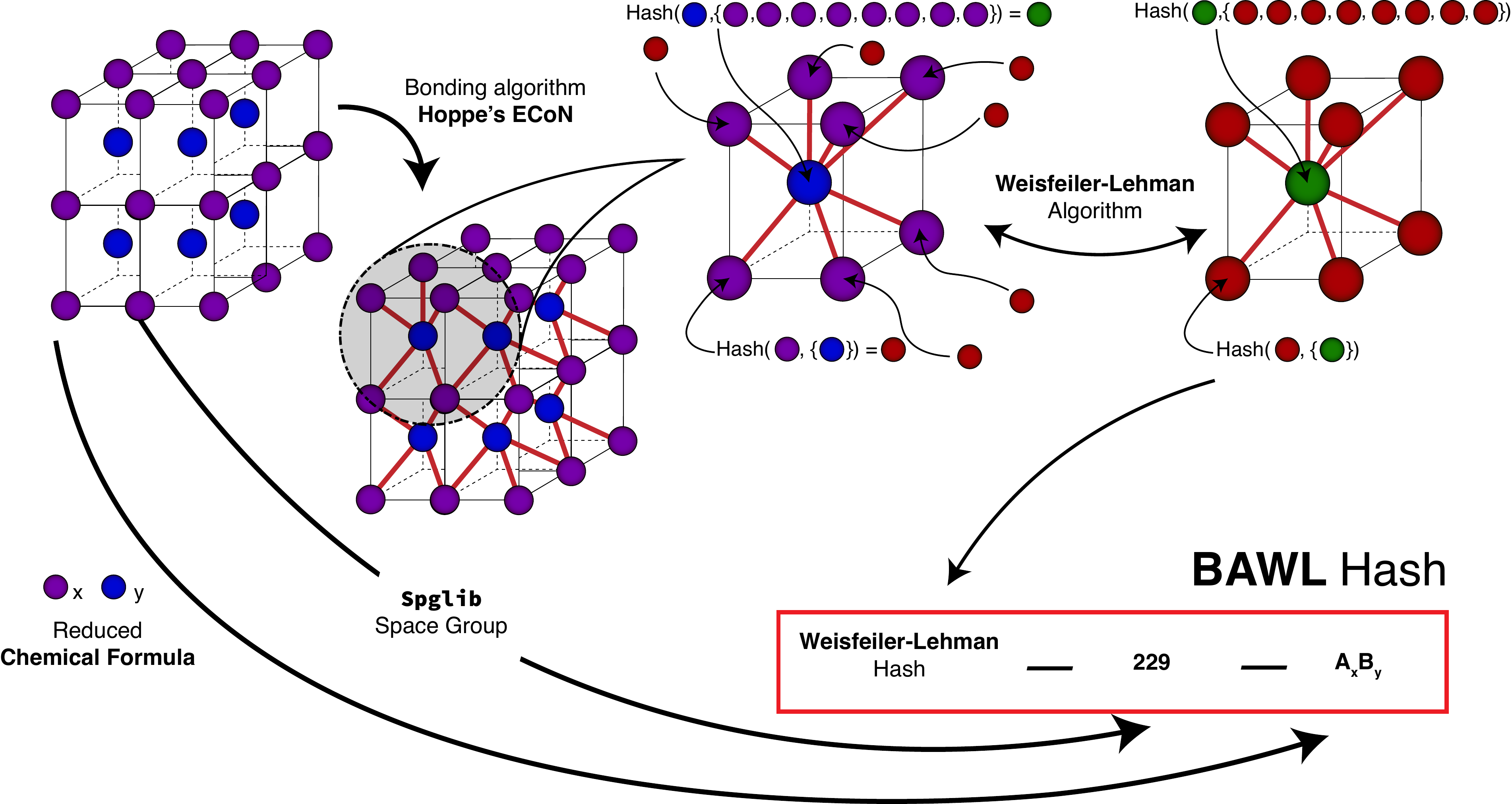}
    \caption{Illustration of the BAWL hashing method. A graph representation of the crystal structure is constructed using the ECoN algorithm. The graph is hashed using Weisfeiller-Lehman. This hash is concatenated with the spacegroup info from Spglib and the reduced chemical formula}
    \label{fig:fingerprint_schematic}
\end{figure}

In order to de-duplicate \textit{LeMat-Bulk}, meaning to identify structurally equivalent materials across different databases, we modify an existing fingerprint technique by Ongari et al.~\citep{ongari2022data} which constructs a bonding graph of a chemical structure, and uses Weisfeiller-Lehman to transform it into a hash. In our modification, to construct the bonding graph, the input structure is initially reduced to its Niggli unit cell using SPGLib~\citep{spglib}. Subsequently, we construct a bonding graph using Effective Coordination Number (ECoN)~\citep{hoppe1979effective} bonding algorithm as implemented in Pymatgen. This bonding graph serves as the basis to compute a Weisfeiler-Lehman (WL) hash. We encode element species as the initial node labels. For WL, we utilize 100 neighbor aggregations to allow for very large atomic system, though this number could very well be reduced, as most unit cells are quite small~\citep{shervashidze2011weisfeiler}. Additionally, in our modification, we construct the final fingerprint by concatenating the WL hash with the unit cell’s space group number and reduced composition, ensuring that both structural and compositional information are encoded. Combining structural and chemical information has previously been deemed important for hashing materials~\citep{yang2014proposed}. A visual depiction of the hash can be found in Fig. \ref{fig:fingerprint_schematic}. Additionally, case studies of identified duplicates using BAWL can be found in the SI for further understanding of this fingerprint technique. We also propose a shortened version of the hash without any space group number (`\textit{Short-BAWL}'). This modification accounts for known limitations in SPGLib-based symmetry detection (sensitivity to tolerance hyperparameters, issues around handling non-conventional unit cell structures, issues processing disordered structures)~\citep{shinohara2023algorithms, spglib}.  

Ultimately, we used the full \textit{BAWL} method to compute fingerprints for all the entries of the \textit{LeMat-Bulk} dataset (i.e., Materials Project, OQMD, and Alexandria). Based on these hashes, we cross-referenced structures—entries with the same hash were considered duplicates—thus establishing links between and within databases. We created a de-duplicated set (\textit{LeMat-Bulk Unique}) in which only the material with the lowest energy among duplicates was kept.

As an additional assessment for structural similarity, we utilized a geometric Graph Neural Network \citep{duval2023hitchhikers}, specifically we compared the untrained EquiformerV2 architecture \citep{liao2023equiformerv2}. We chose an untrained model to compare graph-embedding methods without existing bias on certain materials databases. We utilize EquiformerV2 atom graph embeddings on the energy head to create a vectorized material fingerprint. We then use a cosine similarity function to calculate the distance between pairs of embeddings, establishing this as a similarity score between pairwise structures. We refer to it as \textit{Eqv2-sim} throughout this paper. Analogous to using a hashing function for identifying equivalent materials, we can determine material equivalence using EquiformerV2 graph embeddings between two structures by applying a threshold to their cosine distances. However, this method relies on pairwise comparisons, which constrains computational efficiency.

\subsection{Evaluation framework for fingerprint methods on inorganic crystals}
\label{sec:evaluation-method}

In the following section, we present our evaluation methodology, which tackles: (i) de-duplication, (ii) the robustness of our fingerprint method under different structural perturbations (iii) disordered materials.

\subsubsection{Benchmarking BAWL on LeMat-Bulk}

To establish if entries with matching fingerprints were indeed similar in LeMat-Bulk, we compared energies between compatible entries with identical fingerprints, and then calculated the distribution of energy differences for each matched entry. We specifically analyzed cases with large energy discrepancies exceeding 0.250 eV/atom, a common heuristic for metastability~\citep{aykol2018thermodynamic}. More precisely, we selected two sets of structures for analysis: (1) 35 structure pairs exhibiting energy differences exceeding 5 eV and (2) 150 structure pairs (with same hash) identified using EqV2-sim to have the most dissimilar structures. For each set, we performed DFT relaxations on the structures to evaluate whether large energy differences could be attributed to un-relaxed configurations. Energy and density differences were computed before and after relaxation, allowing us to quantify the extent to which structural relaxation mitigates discrepancies. 

For performing high-throughput DFT we leveraged \texttt{FireWorks}~\citep{jain2015fireworks}, \texttt{atomate2}~\citep{ganose_atomate2_2024}, \texttt{jobflow}~\citep{rosen2024jobflow}, \texttt{custodian}, and \texttt{pymatgen}~\citep{ong2013python} Python packages. As part of the workflow, density functional theory (DFT) calculations were performed using the Vienna Ab-initio Software Package (VASP) \citep{kresse1996efficient, kresse1999ultrasoft} using the Perdew-Burke-Ernzerhof (PBE)\citep{perdew1996generalized} generalized gradient approximation functional with a plane wave basis set with a typical cutoff energy of 450eV and projector augmented wave (PAW) pseudopotentials\citep{blochl1994projector,kresse1999ultrasoft}. Calculations were performed using the Grimme D3 dispersion correction method\citep{grimme2010consistent}. A more stringent energy convergence criterion of \SI{5e-7}{\eV}  was used, and the self-consistent electronic convergence criterion was set to \SI{2e-7}{\eV}. All cell parameters were allowed to relax. A 1,000 reciprocal k-points density was chosen. For most calculations, symmetries were turned off for VASP. Custodian was used to fix calculations with poor convergence on the fly, while keeping the stringent convergence criteria. For systems tested with large amount of noise we applied the conjugate gradient optimization method with a larger step size of 0.5.

\subsubsection{Benchmark fingerprints' sensitivity and robustness}

To systematically evaluate the sensitivity of our proposed fingerprinting method, we designed a set of benchmarks assessing the threshold of its ability to distinguish crystal structures. The idea being that a robust fingerprint should be invariant to translations and symmetry operations that preserve identity of the material, while being sensitive to meaningful structural differences. However, defining the threshold at which a structure should be considered "novel" remains ambiguous. We postulate in this paper that small perturbations in lattice vectors or atomic positions may not significantly impact an inorganic crystalline material's identity, making it necessary to quantify fingerprint sensitivity to structural noise. However, required sensitivity could be dependent on application or material needs. For example, more flexible materials such as MOFs may necessitate larger sensitivity~\citep{ongari2022data}.

For this experiment, we thus sampled 100 materials from \textit{LeMat-Bulk} and applied varying level of random noise to either their atomic coordinates or their lattice vectors. We then calculated the resulting fingerprint and determined a threshold at which the altered structure no longer matched its original fingerprint. We defined the \textit{sensitivity} as the noise at which the hashing methods are no longer able to match the original structure to its perturbed counterpart. 

Similarly, we measured their invariance to symmetry and lattice translation operations, ensuring consistent fingerprinting for structures that remain equivalent under these transformations. We used SPGLib as implemented in Pymatgen to detect affine transformations that could be applied to a given material, and calculated a fingerprint for all possible affine transformations for each given material.

We compare the performance of the BAWL and Short-BAWL fingerprint methods against several other structure comparison techniques. A comparison of previous work on structure similarity methods and structure fingerprint methods can be found in the SI. However, none of the techniques we compared have been thoroughly benchmarked across the tests described in this paper. In comparing structure matching algorithm, we divided methods into either fingerprint techniques, if they could return string-like information about the structure, or similarity techniques, if they returned numerical information about the structures, or if a threshold was needed to characterize whether two structures were equivalent. In total, we compared four similarity methods, including two Structure Matching algorithm - \textit{Pymatgen's StructureMatcher} and \textit{Mattergen's Disordered Structure Matcher} - as well as two numerical methods - PDD\citep{widdowson2022resolving} and eqV2-sim. For PDD, we utilized Wasserstein distance to match the structures. For Pymatgen's and Mattergen's Structure Matchers we utilized all default parameters. For both PDD and eqV2 we set a numerical threshold of 0.01 for defining equivalence. In setting these threshold, we attempted to match the number of unique structures detected by both BAWL and these methods. These methods use a pairwise combinatorial approach to identify duplicate structures rather than hashing. For fingerprints, we assessed  SLICES \citep{xiao2023invertible}, and CLOUD \citep{xu2024cloud}, using our curated set of benchmarks. 
Similarly, we test their invariance to lattice translations and symmetry operations.

This evaluation framework provides a comprehensive assessment of fingerprinting algorithms, measuring their balance between stability and sensitivity—ensuring that they remain robust to minor perturbations while correctly distinguishing truly novel materials. Additionally, we measured how long it would take to assess whether a set of structures using each fingerprint and similarity methods take, and extrapolated this for de-duplicating across \textit{LeMat-Bulk}.

\subsubsection{Disordered materials}

Lastly, we proposed a method to evaluate the ability of hashing methods to handle disordered structures. We constructed a dataset of disordered materials, for which fingerprint and similarity algorithms can be benchmarked against. To construct the database, we employed the Supercell software package \citep{okhotnikov2016supercell}, which generates substitutionally disordered structures based on user-defined inputs and symmetry constraints. The dataset includes supercell representations of disordered materials, curated from the Supercell study itself. The dataset spans various disorder types, including substitutional disorder (\ce{Sn_{0.5}Pb_{0.5}Te}, \ce{MnCaCO3}, \ce{Ti_{1-x}Mg_{x}N}), proton disorder in Ice Ih, and cationic site disorder (\ce{Ca2Al2SiO7}, CZTSe, \ce{FeSbO4}, \ce{MgAlFeO4}, PZT, \ce{Rb}-PST-1, \ce{SrSiAlO_{x}}). These materials exhibit disorder affecting thermoelectric, ferroelectric, catalytic, optical, and mechanical properties \citep{parker2013high, kuo2005effect, florian2013beyond, dun2014cu2znsnsxo4, grau2006electronic, nili2022magnetic, mucci1988manganese, bell2006factors, cadars2017modeling, srirangam2014probing, benna1995si, wang2021tunable}.

To evaluate fingerprinting and similarity methods, we compared the BAWL and Short-BAWL fingerprint against SLICES, and CLOUD, as well as Pymatgen's and Mattergen's structure matchers, Eqv2-sim and PDD, measuring their effectiveness in matching disordered structures.

\subsection{Comparing trends across functionals}

After having established confidence in our hashing technique, we investigate how key trends vary across different DFT approximations, providing insights into the impact of functional choices on material properties. More precisely, we utilized BAWL to systematically compare material properties across functionals, matching structures across functionals. When duplicates were present, we selected the structure with the smallest energy per atom for each functional. We then took the lowest energy entries' corresponding energy, magnetization, and fermi energy. We utilized previously described method to compute formation energies. We compared all three functionals: PBE, PBESol, and SCAN.

\subsection{Software availability}
The code for database integration can be found here: \url{https://github.com/LeMaterial/lematerial-fetcher}. The code for structure fingerprint and similarity benchmarks can be found here: \url{https://github.com/LeMaterial/lematerial-hasher}

\section{Results}
\subsection{LeMat-Bulk database}
\begin{figure}
\centering
\includegraphics[height=10cm]{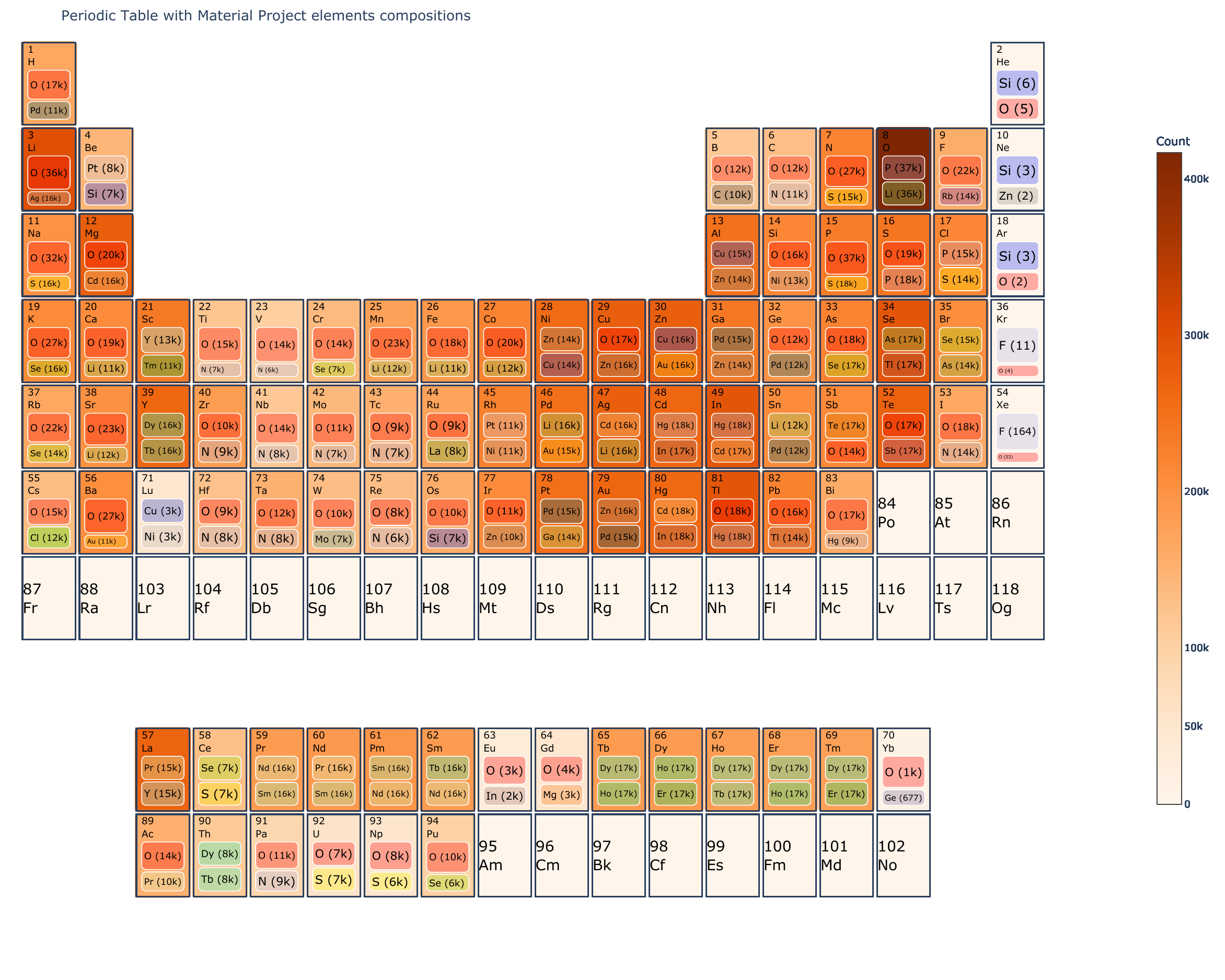}%
\caption{2D block chart demonstrating chemical bias in LeMat-Bulk. Less bias is present in LeMat-Bulk compared to any single database.\label{fig:chemicalbias}}
\end{figure}

The resulting \textit{LeMat-Bulk} dataset comprises 5.34 million materials, including contributions from the Materials Project (146,000 entries), OQMD (568,000 entries), and Alexandria (4.62 million entries). This marks the first public release of a unified dataset integrating data from these three major repositories. Within the dataset, 448,000 materials were calculated using the PBESol functional, and 423,000 were calculated using the SCAN functional. Additionally, 520,000 entries that were incompatible with the dataset's calculation parameters were reformatted for the community to benefit from a single database. The database unites structural properties via the Optimade format, along with energy, total magnetization, stress tensor, forces, and magnetic moments (Table. \ref{tab:lemat}).

Notably, \textit{LeMat-Bulk} addresses some of the chemical biases that are inevitable in any individual database. While the Materials Project provides extensive coverage of transition metal oxides and battery-relevant materials (e.g., Li, O, P systems), integrating data from OQMD and Alexandria broadens the overall scope. For example, in columns 10–12 of the periodic table, \textit{LeMat-Bulk} captures a wider range of bimetallic compounds: whereas the Materials Project includes roughly 4,000 Ni,O-containing compounds out of 10,000 total Ni-containing compounds, \textit{LeMat-Bulk} incorporates 14,000 Ni,Zn and 14,000 Ni,Cu. This more diverse coverage extends to rare-earth elements as well; for instance, the Materials Project features about 3,000 La,O-containing compounds, while \textit{LeMat-Bulk} includes 15,000 La,Pr-containing compounds, highlighting an expanded focus on multi-rare-earth systems. Of course, no dataset is entirely free of bias: oxygen-containing compounds remain overrepresented, and the large contribution from Alexandria shifts \textit{LeMat-Bulk} toward that repository’s focus areas. Even so, the overall degree of bias is reduced compared to using a single database. Such biases, as documented in earlier work~\citep{zhang2023mitigating,horton2021promises}, can limit robust machine learning model development~\citep{zhang2023mitigating,trezza2024classification,decost2020scientific}. By mitigating some of them, \textit{LeMat-Bulk} offers a more balanced foundation for machine learning applications and advances our understanding of materials chemistry. A visual of the chemical bias of \textit{LeMat-Bulk} can be found in Fig. \ref{fig:chemicalbias}, and the bias of its constituent database can be found in the SI in Fig. \ref{fig:bias-mp},\ref{fig:bias-alexandria}, and \ref{fig:bias-oqmd}.

\begin{figure}
\centering
  \includegraphics[height=9cm]{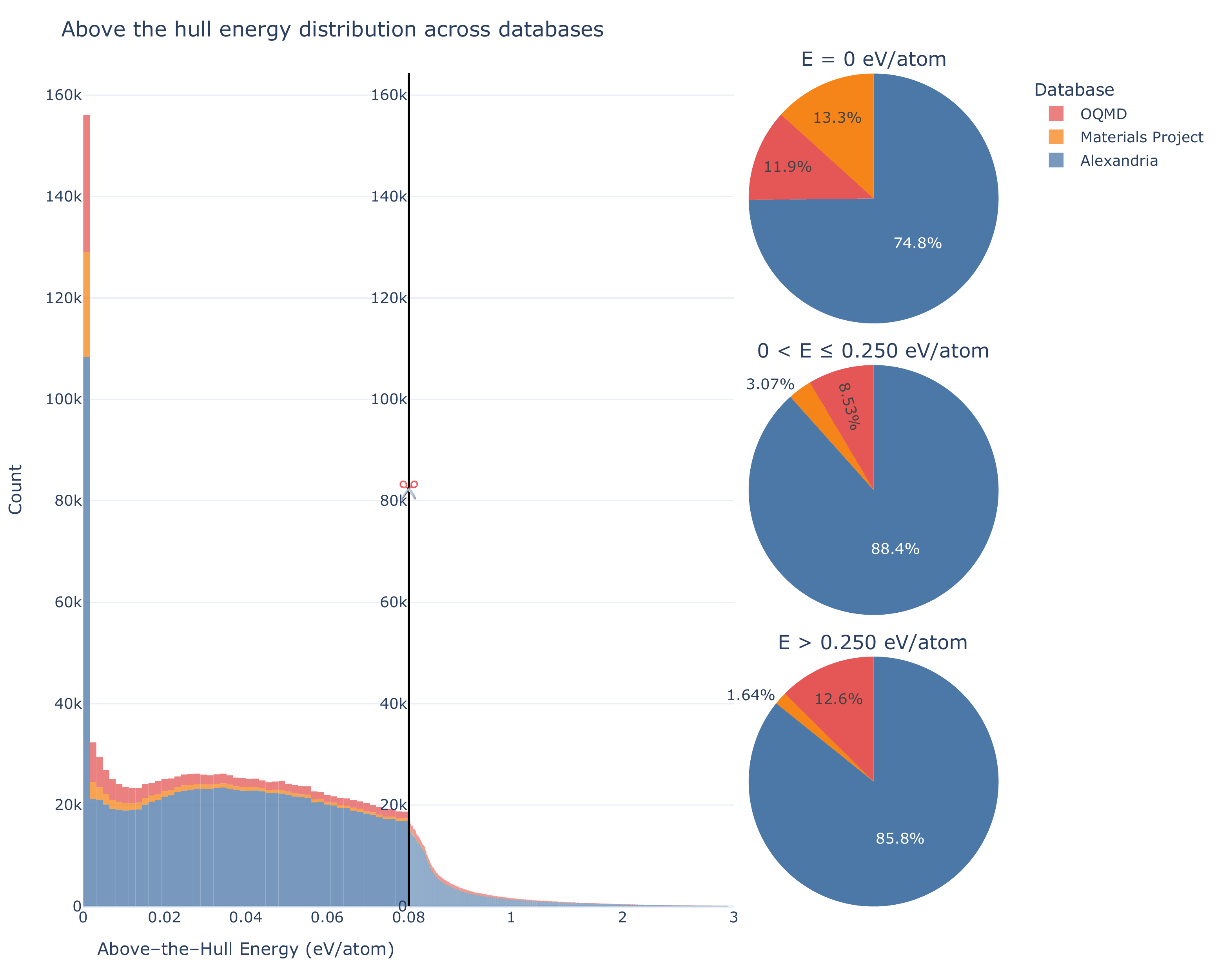}
  \caption{Distribution of above the hull materials across all materials in LeMat-Bulk by database of origin. Pie chart showcase percent of materials originating from each database that are on the hull, potentially metastable and unstable.}
  \label{fig:distribution}
\label{fig:test}
\end{figure}

Additionally, the inclusion of the Alexandria dataset reduces bias towards thermodynamically stable materials (which are `on the hull'), offering more comprehensive representations of metastable materials (a typical heuristic of 250 meV/atom is used to define metastability). This allows models trained on this dataset to learn from materials that may be thermodynamically accessible but are often underrepresented in common quantum chemistry databases. By combining these three databases, we found 108,000 entries that are on the hull, with 81,000 coming from Alexandria, 14,000 from Materials Project, and 13,000 from OQMD (Fig. \ref{fig:distribution}). This integration demonstrates that merging these databases enables a richer description of material stability across compositional spaces.

LeMat-Bulk includes a subset of material entries that are not compatible. These are materials we consider incompatible for combination due to their calculation parameters, but we have formatted them for others to use. When comparing materials with the same fingerprints and their corresponding range of energies in the PBE subset of compatible data points, only 1\% of matching entries exceeded an energy difference of 0.250 eV/atom, with all data points remaining below 10 eV/atom. In contrast, when comparing materials with the same fingerprints between the compatible and non-compatible entries, over 10\% of the resulting entries exhibited energy differences greater than 0.250 eV/atom, with some mismatches as high as 50 eV. This implies that duplicate materials calculated with parameters we deemed incompatible yielded significantly different structural energies. This reinforces the importance of using compatible calculation parameters when combining datasets prior to training models such as force fields.

We also observed that a significant proportion of matching compounds within only the non-compatible subset exhibited a large range of energy differences, suggesting that even within this subset, there are entries that may not be mutually compatible. However, most non-compatible records come from OQMD. Interestingly, the number of matching materials with energy differences greater than 3 eV was similar between the non-compatible and compatible subsets, despite the non-compatible subset being an order of magnitude smaller than the compatible dataset.

These findings support the validity of our alignment choices for calculation parameters. Additionally, we found a substantial proportion of matching structures with similar energy between the compatible and non-compatible subsets, suggesting that our parameter filtering approach may be conservative regarding certain parameters. Likely, the key factors are spin polarization rather than pseudopotential or Hubbard U. In fact, when we look at matched entries between the compatible and non-compatible subset by fingerprint, we find that out of those that have small energy difference, below 0.01eV, 73\% do not have any d- or f- electrons. The percentage drops to 40\% for differences above 5eV. Of course, in the non d- or f- electron-containing calculation there are still pseudopotential differences.

\subsection{Evaluation of hashing methods}

In this section, we provide the results of the evaluation methodology presented in Section \ref{sec:evaluation-method}, targeting de-duplication, sensitivity to structural perturbations and disordered structures.

\subsubsection{De-duplication of LeMat-Bulk with BAWL}
Using the BAWL fingerprint method, we identified 50,580 entries shared between Alexandria and Materials Project. Extending the analysis with the Short-BAWL fingerprint, we increased the match count to 54,483 structures. Accounting for intra-database duplicates within the Materials Project (approximately 4,700) and excluding materials due to parameter incompatibility (approximately 13,000), we estimate that over 40\% of the Materials Project entries were successfully matched in Alexandria. Furthermore,  we identified approximately 26,000 OQMD structures in Alexandria and 1,800 OQMD structures within the Materials Project. Our method, in terms of application for database matching operates with higher sensitivity. Nevertheless, the ability to match structures across databases, allows users to access properties from one database not calculated in another for the same material.

To test the robustness of our hash in de-duplicating materials databases, we analyzed entries identified as duplicates. We found over 35 structures with identical BAWL fingerprints yet exhibiting substantial energy differences in \textit{LeMat-Bulk}. Initially, these structures showed large energy and density variations. After DFT relaxation, both energy and density differences significantly decreased. During this analysis, we identified nine structures exhibiting unusually large initial energy differences due to positive total energies in their corresponding OQMD entries, likely resulting from improper calculation parameters. This highlights the utility of the BAWL fingerprint approach in detecting calculation inconsistencies.

In a complementary analysis, we evaluated 150 structure pairs with identical BAWL fingerprints but highly dissimilar EqV2-sim embeddings. Detailed analyses of energies and densities before and after DFT relaxation are summarized in Table \ref{tab:results_summary}, with corresponding plots provided in the supplementary information (Fig. \ref{fig:input_output_density_most_dissimilar_energies}).

\renewcommand{\arraystretch}{1}
\begin{table}[htbp]
    \centering
    \caption{Summary of DFT-based duplicate analysis using BAWL fingerprints}
    \label{tab:results_summary}
    \resizebox{\textwidth}{!}{
    \begin{tabular}{>{\raggedright\arraybackslash}p{0.18\linewidth} cc cc}
        \toprule
        \multirow{2}{*}{\textbf{Analysis}} 
        & \multicolumn{2}{c}{\textbf{Mean energy difference (eV)}} 
        & \multicolumn{2}{c}{\textbf{Mean density difference (g/cm$^3$)}} \\ 
        \cmidrule(lr){2-3} \cmidrule(lr){4-5}
        & Before relaxation & After relaxation 
        & Before relaxation & After relaxation \\
        \midrule
        Large energy difference
            & \shortstack{10 \\(5.65--46.00)} & \shortstack{0.07 \\($8 \times 10^{-8}$--2.18)}
            & \shortstack{0.55 \\(0.075--3.98)} & \shortstack{0.17 \\($7 \times 10^{-7}$--2.75)} \\
        \addlinespace
        Dissimilar EqV2-sim
            & \shortstack{0.21 \\(0.13--0.24)} & \shortstack{0.03 \\(0.026--0.033)}
            & \shortstack{1.71 \\(1.41--1.85)} & \shortstack{0.10 \\(0.02--0.16)} \\
        \bottomrule
    \end{tabular}
    }
\end{table}

These findings demonstrate that many large energy and density discrepancies between structures with identical BAWL fingerprints can be likely attributed to un-relaxed structures rather than fundamental differences in materials. In fact, we detected many materials with large norms of forces, primarily in Materials Project and OQMD (SI Fig. \ref{fig:large_force} and Fig. \ref{fig:max_force}). The ability of our fingerprinting approach to detect such cases further validates its robustness.

\subsubsection{Sensitivity to perturbations against other hashing algorithm}
\begin{figure}
\includegraphics[height=8cm]{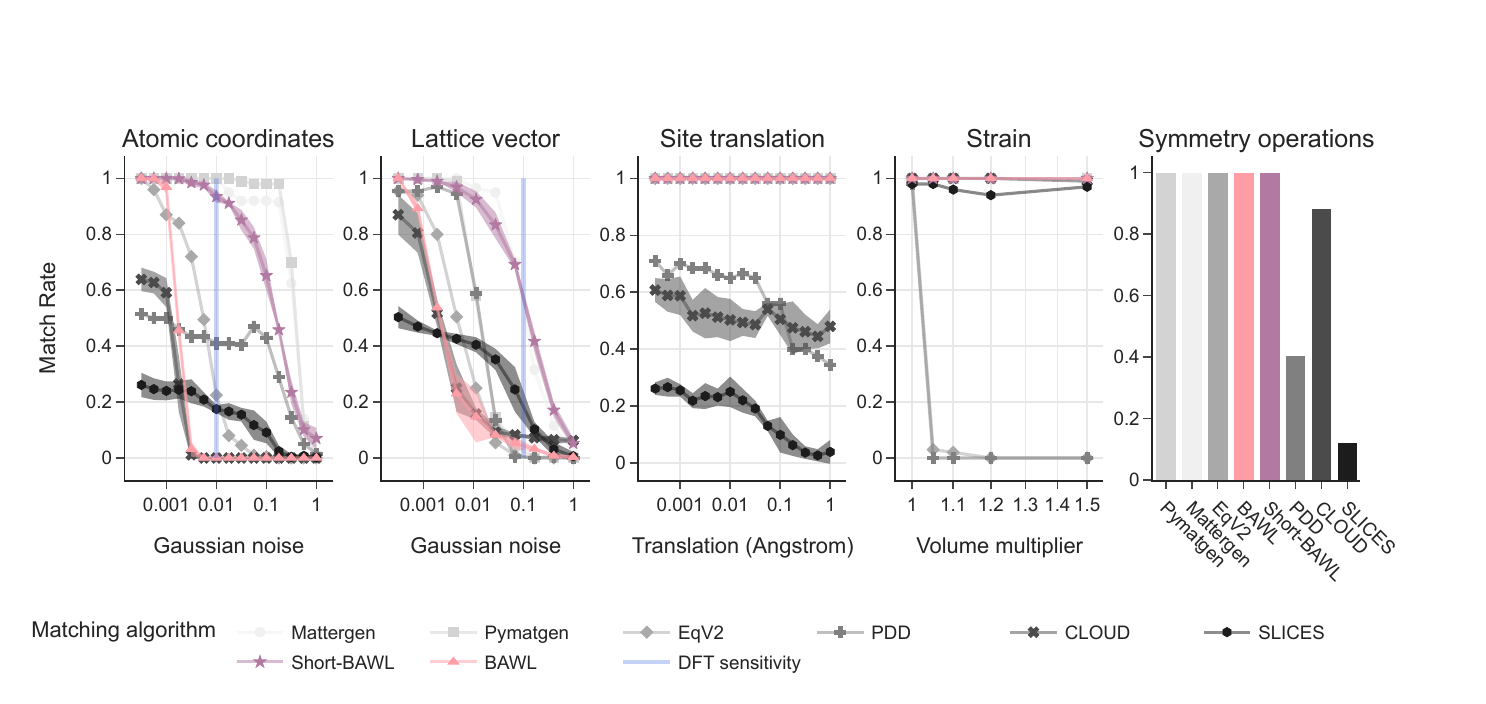}%
\caption{Benchmark results of various fingerprint and similarity methods across noise on atomic coordinates, noise on lattice vectors, translation, strain and crystal symmetry operations. Inclusion of DFT sensitivity based on DFT relaxation tests. Thickness around lines indicates areas of higher variability across the methods tested.}
\label{fig:benchmark_results}
\end{figure}

To further benchmark the hashing method, we tested its sensitivity across various structure perturbations and compared it to other hashing and similarity algorithms. Full results of this can be found in Fig. \ref{fig:benchmark_results}.

The BAWL fingerprint exhibited a sharp robustness cut off at approximately 0.002 gaussian noise on atomic fractional coordinates, largely due to its reliance on SPGLib's symmetry detection. In contrast, the Short-BAWL fingerprint demonstrated significantly improved tolerance, with more gradual sensitivity drop, with a cutoff of gaussian noise on atomic coordinates of 0.2. SLICES and its modified version, CLOUD, exhibited even higher sensitivity: being quite sensitive beyond 0.001 noise. Despite this, SLICES retained some robustness at higher noise compared to the BAWL fingerprint. In our DFT experiments, we find the sensitivity (with respect to our DFT parameters) to be around 0.01 for noise on atomic coordinates. Out of the fingerprint methods, only Short-BAWL fingerprint has a sensitivity above that of DFT, however, much higher. When comparing similarity methods, Pymatgen and Mattergen's structure matchers have very low sensitivity at a noise of 0.8 each, with Pymatgen having slighly lower sensitivity. EqV2-sim exhibited robustness similar to DFT, with a slightly lower sensitivity than BAWL. PDD was the most sensitive out of the similarity and hashing methods, with a sharp cutoff at 0.003 atomic coordinate noise. 

For noise on lattice vectors, the Short-BAWL fingerprint and Mattergen's StructureMatcher were the most robust, having very similar sensitivity curves, with a sensitivity of around 0.1. This was followed by Pymatgen's StructureMatcher and PDD which have very similar profiles, then EqV2-sim, BAWL, and CLOUD, with respective thresholds (at 50\% matching) of 0.02, 0.01, 0.001, 0.001, and 0.0005. The Short-BAWL fingerprint was the only fingerprint capable of maintaining structure matches under minimal lattice parameter noise, underscoring the advantages of avoiding symmetry-based encoding when benchmarked for perturbation tolerance. In our DFT experiments, we find the sensitivity (with respect to our DFT parameters) to be around 0.1 for noise on lattice vectors. In our fingerprint experimentation, only Short-BAWL had a similar sensitivity as DFT, with all other fingerprint being more sensitive.

Both the BAWL and Short-BAWL fingerprints exhibited complete invariance to site translations. All other fingerprints performed poorly. SLICES failed to match structures in more than 50\% of cases under site translations. For CLOUD and SLICES, this is likely due to their reliance on Wyckoff position re-labeling. For structure matchers, both Mattergen's, Pymatgen's, and EquiformerV2 embeddings performed perfectly under lattice translation, matching BAWL and Short-BAWL. However, PDD failed under most small levels of site translations.

Only PDD and EquiformerV2 had significant sensitivity under isometric lattice strain. All other fingerprints did not show much sensitivity, maintaining consistent fingerprints even at extreme strain levels of 50\%.

For symmetry operations, BAWL and Short-BAWL were the only fingerprint methods to achieve 100\% accuracy in matching structures under symmetry operations, indicating strong invariance to these transformations across methods. All similarity methods except PDD (EqV2-sim, Pymatgen and Mattergen) also had 100\% success. However, none of the other fingerprint methods (CLOUD, SLICES) had success in generating matching fingerprints under symmetry operations. All fingerprint and similarity benchmark to structure perturbations can be found in Fig. \ref{fig:benchmark_results}.

CLOUD and SLICES, heavily reliant on symmetry detection and Wyckoff labeling, were disproportionately sensitive to minor structural perturbations, amplifying their failure rates under noise and translations. The WL methods emerged as the most robust fingerprint methods, successfully handling all invariances of crystal systems: lattice translations and symmetry operations. The Short-BAWL was also the least sensitive to atomic position displacements and lattice vector noise. These findings highlight WL methods as the most robust and reliable fingerprinting methods for detecting duplicate materials and robustly fingerprinting crystal structures.

For structure matching, PDD performed the worse across all test cases. While structure matching algorithms (except PDD) performed well under these varying transformations (although both Pymatgen and Mattergen are arguably too insensitive to lattice noise), their combinationarial approach limits their use for database de-duplication in large database settings. While fingerprint method scale linearly in deduplicating, combinatorial approach scale quadratically. On LeMat-Bulk, deduplicating with Mattergen's structure matching algorithm would have taken over a year of compute compared to minutes for BAWL and Short-BAWL (Fig. \ref{fig:pairwise_extrapolate_time} and Fig. \ref{fig:pairwise_time}).

\subsubsection{Disordered structures}
\begin{table}[h]
    \centering
    \begin{tabular}{lcc}
        \toprule
        \textbf{Method} & Matching rate across compositions (\%) \\
        \midrule
        \textbf{BAWL} &  10 $\pm$ 32 \\
        \textbf{Short-BAWL} &  33 $\pm$ 33 \\
        \textbf{EqV2-sim} &  61 $\pm$ 38 \\
        \textbf{PDD} &  89 $\pm$ 41 \\
        \textbf{Pymatgen} &  0.0 $\pm$ 1 \\
        \textbf{CLOUD} &  26 $\pm$ 33 \\
        \textbf{SLICES} &  0.0 $\pm$ 0.00 \\
        \textbf{Mattergen} &  87 $\pm$ 35 \\
        \bottomrule
    \end{tabular}
    \caption{Comparison of different methods ability to properly match disordered materials. Across each composition, we report how many of each structures were able to be properly matched. We then aggregate these results over all chemical formulas and report the mean success rate and standard deviation. Additionally, we tested false positives by attempting to match one composition to another. No method returned any false positives. }
    \label{tab:comparison}
\end{table}

In testing the varying similarity and fingerprint methods on the disordered structures dataset, we found that PDD was the most successful overall method, scoring 100\% across multiple materials systems (\ce{Ca2Al2SiO7}, \ce{Fe2Cu4Sn2Se5S3},\ce{FeSbO4}, \ce{MgAlFeO4}, \ce{Sn_{0.5}Pb_{0.5}Te}, \ce{ZrTiPb2O6}) failing completely with water, and failing partially with \ce{SrSiAlO_{x}}. Meanwhile, the Short-BAWL fingerprint was the most successful out of all fingerprint methods. However, it only succeeded in one case, \ce{FeSbO4}, and failed with low rates in other (water scoring 60\%, \ce{Fe2Cu4Sn2Se5S3}, scoring 35\%, and others scoring less than 25\%) while completely failing in one case (\ce{MgAlFeO4}). All other fingerprint methods scored worse. SLICES failed at detecting all disorded cases. CLOUD meanwhile only has noticeable success with \ce{SrSiAlO_{x}}, although it fared better with \ce{Fe2Cu4Sn2Se5S3} than Short-BAWL. Besides PDD, for the similarity algorithms, Mattergen performed the best only failing with the water system. EqV2-sim was a close second, scoring 100\% in multiple systems (\ce{Fe2Cu4Sn2Se5S3}, \ce{MgAlFeO4}, \ce{ZrTiPb2O6}) while failing greatly with water. It performed moderately well on other systems. Pymatgen meanwhile failed across the board on all systems. 

We expect the WL with or without symmetry method to fail for two reasons: the graph bond may be different depending on the ordering of the atoms, and the symmetry label might thus be different as well. Some cases would lead to the same bonding graph which thus should have a better success rate than certain other methods. Surprisingly, BAWL fingerprint had the highest success among the water structures, which all other methods failed. This showcases the strength of using bonding algorithm for matching. Symmetry reliant algorithms failed greatly in the disordered test cases. For the BAWL and Short-BAWL, we note that the bonding hash graph could be modified to include for partial occupancy and properly match certain hashes which fill these partial occupancy, however most chemical databases do not retain such information for disordered materials. PDD performed surprisingly well, in part due to the fact that it does not embed composition well hashing its point clouds, rendering it more robust with respect to composition disorder. We expect instead similarity metrics to serve as a better tool in this space, especially equipped with a threshold value to determine material similarity, however we note that the problem with these metrics, like Pymatgen's Structure Matcher is that these are used to determine whether two structures are the same rather than provide a fingerprint which can be looked up. Thus they require to run in an combinatorial approach around all structures, severely limiting their use. A summary of the results of the disordered structure benchmark can be found in Table \ref{tab:comparison}.

Overall, for fingerprint-based algorithms, we found that WL-based algorithms preserved all symmetries and affine transformations of the bulk lattice. While more sensitive to noise in atomic coordinates and lattice vectors than the structure similarity method, it was the least sensitive among the fingerprint methods. Additionally, all fingerprint methods were insensitive to strain. For these reasons, we chose BAWL for both deduplication and comparing trends across functionals.

\subsection{Comparing trends across functionals}

\begin{figure}
\includegraphics[height=10cm]{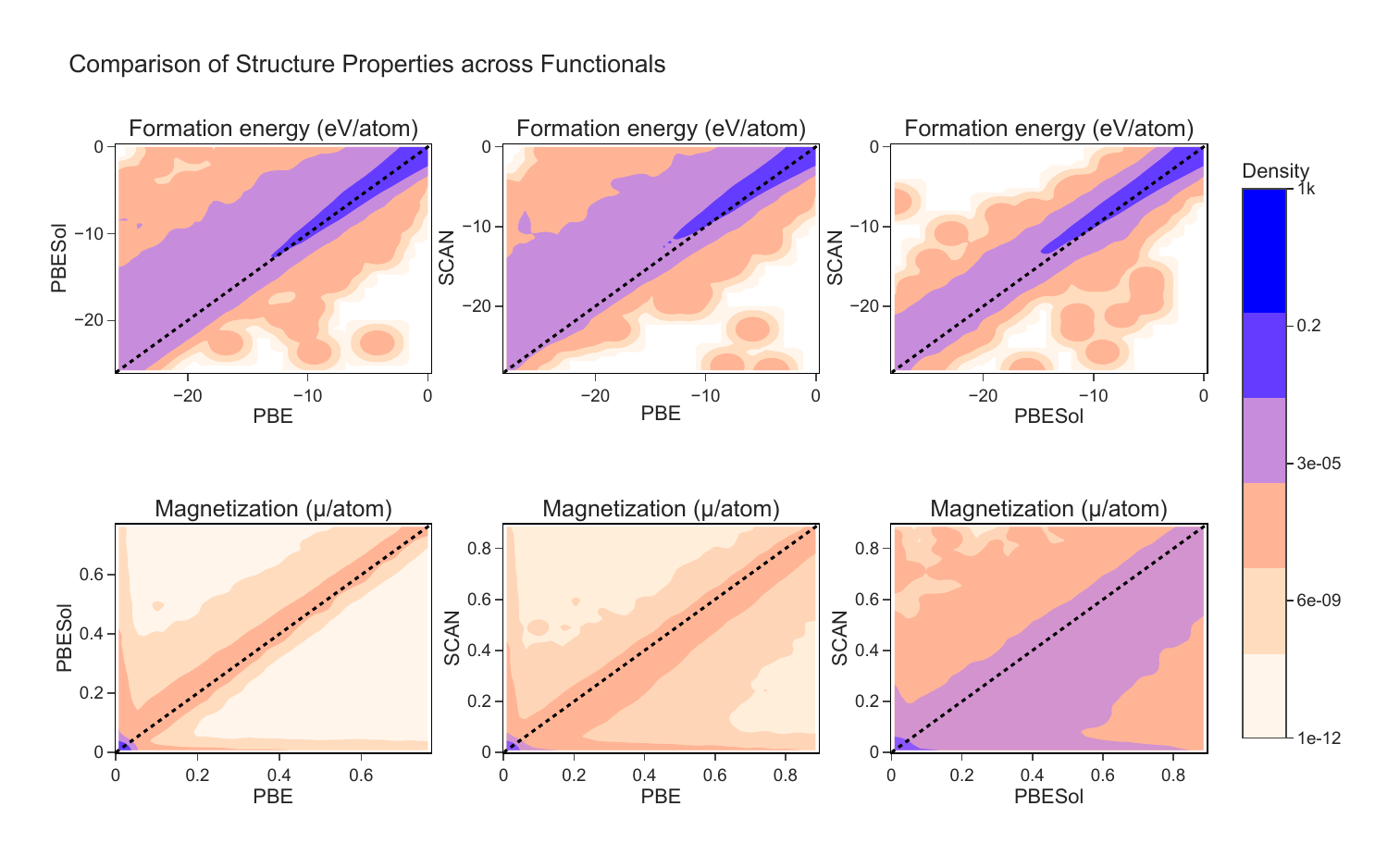}%
\caption{Correlation density heatmap of formation energy and total magnetization normalized per atom across three functionals (PBE, PBESol, and SCAN). Materials were matched across functionals using BAWL fingerprint methods. Where a set of materials with the same fingerprint exist for a given functional, the lowest energy material was used for all properties. For total magnetization the absolutes were used only. The dashed black line indicates the parity line.}
\label{fig:correlation_functional}
\end{figure}

With increased confidence in the hashing algorithm, we now leverage BAWL-computed fingerprints to establish correspondences between materials calculated under different computational settings. This capability is particularly powerful: for example, within the OQMD database, if a given band gap is computed under specific parameters, we can examine how that value changes as a function of Hubbard U, pseudopotentials, k-point density, or even exchange-correlation functional. To rapidly assess the ability of our framework to capture such variations, we performed a comparative analysis of material properties across different functionals within the LeMat-Bulk dataset, focusing on total magnetization and formation energy for PBE, PBEsol, and SCAN. While a comprehensive evaluation of functional-dependent property variations lies beyond the scope of this study, our goal is to demonstrate the feasibility of such an analysis at the database level. As described in the Methods section, we conducted pairwise comparisons of formation energy per atom and total magnetization per atom across all matched materials in LeMat-Bulk-Unique. Although the specific response of each property to functional choice is system-dependent and warrants a more exhaustive investigation, general trends can nonetheless be observed across functionals.

In our broad, chemically agnostic analysis of the impact of exchange-correlation functional choice on bulk material properties within the LeMat-Bulk dataset, we observe that PBE systematically overbinds relative to PBESol, while PBESol exhibits stronger correlation with SCAN, though it also tends to overbind. Prior studies have indicated that SCAN often underbinds alkali metals~\citep{kovacs2019comparative}, and it has been shown to systematically underpredict formation energies for strongly bound compounds relative to PBE, with this trend being less pronounced in the weakly bound regime~\citep{isaacs2018performance}. Furthermore, SCAN has been reported to yield the largest predicted lattice constants for strongly bound materials—consistent with weaker binding—while for weakly bound materials, both SCAN and PBE have been shown to provide comparable lattice volume predictions~\citep{kingsbury2022performance}. In our analysis, we similarly find that SCAN predicts lower formation energies for strongly bound compounds compared to PBE and PBESol. Additionally, we observe a high density of near-parity data points in the weakly bound regime when comparing SCAN to both PBE and PBESol, on par with literature.

In comparing the total absolute magnetization across exchange-correlation functionals for materials in the LeMat-Bulk dataset, we observe that PBE generally and systematically underpredicts magnetization relative to PBESol, while tending to overpredict it relative to SCAN. Additionally, SCAN and PBESol exhibit a stronger correlation in total magnetization compared to the correlation between PBE and SCAN. We find a significant number of cases where one functional predicts finite magnetization while others predict none. Prior studies examining the chemical dependence of magnetization across functionals have shown, for example, that in Fe, Co, and Ni, SCAN produces values intermediate between those obtained with PBE and PBE+U, with PBE+U generally providing better agreement with experiment~\citep{fu2019density}. Other studies have reported that SCAN tends to overestimate magnetization in transition metals~\citep{mejia2019analysis}, while yet another study has found that SCAN can offer systematic improvements in magnetization predictions~\citep{isaacs2018performance}. These findings suggest that a composition-dependent analysis of magnetization trends may be warranted, though such an investigation lies beyond the scope of this work. Instead, our aim is to demonstrate that, with a robust fingerprinting and hashing framework, it is possible to establish meaningful correspondences between quantum chemistry calculations performed under varying computational parameters.

The observance of clearly defined trends across these two properties, demonstrate that the material fingerprint method effectively matches structures across datasets and functionals, enabling a robust comparison of energy, and magnetization. Our findings largely aligns with existing literature. Our findings provide further insight into the performance and limitations of these exchange-correlation functionals, with implications for selecting appropriate functionals for specific material systems.

\section{Conclusion}

In this study, we addressed critical challenges in quantum chemistry databases, including data redundancy, inconsistencies in computational parameters, and the lack of standardized benchmarks for material novelty. By integrating data from the Materials Project, OQMD, and Alexandria databases, we developed the \textit{LeMat-Bulk} dataset, comprising over 5.3 million materials and representing the largest publicly available dataset to date with PBE, PBESol, and SCAN exchange-correlation functional calculations. This integration mitigates the chemical and structural biases present in individual databases, enhances representation of metastable materials, and provides a foundation for broader applications in machine learning and materials discovery.

To identify duplicate and similar materials, we proposed a new fingerprinting method, BAWL, and its variant Short-BAWL, based on bonding graphs and Weisfeiler-Lehman hashing. Our benchmark across various perturbations -- atomic coordinate noise, lattice strain, lattice translations, and symmetry operations -- demonstrated that both BAWL and Short-BAWL outperform existing fingerprinting techniques (e.g., SLICES, and CLOUD) in robustness and speed. In the case of disordered systems, Short-BAWL emerged as the most reliable fingerprinting approach, while Mattergen’s StructureMatcher performed best among similarity metrics. Nonetheless, our results highlighted open challenges, including the need to encode compositional disorder more effectively and to refine fingerprinting schemes for highly complex structures. Additionally, our fingerprinting approach proved effective in identifying trends across functional spaces, allowing us to reconcile discrepancies in energy, and magnetization properties across PBE, PBESol, and SCAN functionals.

Despite these advancements, challenges remain. Symmetry-based encoding methods like BAWL exhibited sensitivity to perturbations and failed for highly disordered systems. Similarity metrics (e.g., EquiformerV2 embeddings) achieved high accuracy in certain benchmarks for detecting structural equivalences but their reliance on computationally expensive combinatorial approaches limits scalability. Future work should focus on developing more efficient similarity metrics and hashing techniques for robust handling of complex disordered systems and expanding database interoperability.

Overall, this work represents a significant step toward standardized, integrated, and scalable approaches for quantum chemistry database management, facilitating reproducibility, accelerating materials discovery, and advancing generative materials 
science.

\section*{Acknowledgments}

We acknowledge HuggingFace and their team including Thomas Wolff for providing compute for calculating Bader charges and hosting of the database, and Leandro Von Werra for calculating Bader charges. We acknowledge fruitful discussions from Zack Ulissi, and Matt Horton. EF and PS acknowledge support from the NCCR Catalysis (grant number 225147), a National Centre of Competence in Research funded by the Swiss National Science Foundation.

\bibliographystyle{unsrtnat}
\bibliography{references}

%%%%%%%%%%%%%%%%%%%%%%%%%%%%%%%%%%%%%%%%%%%%%%%%%%%%%%%%%%%%

\appendix

\section{Case studies to understand BAWL hashing}
\subsection{Case study: specific hashing example for pair of duplicate found in Alexandria}

\begin{figure}[htbp]
    \centering
    % First figure
    \begin{subfigure}[b]{0.3\textwidth}
        \includegraphics[width=\textwidth]{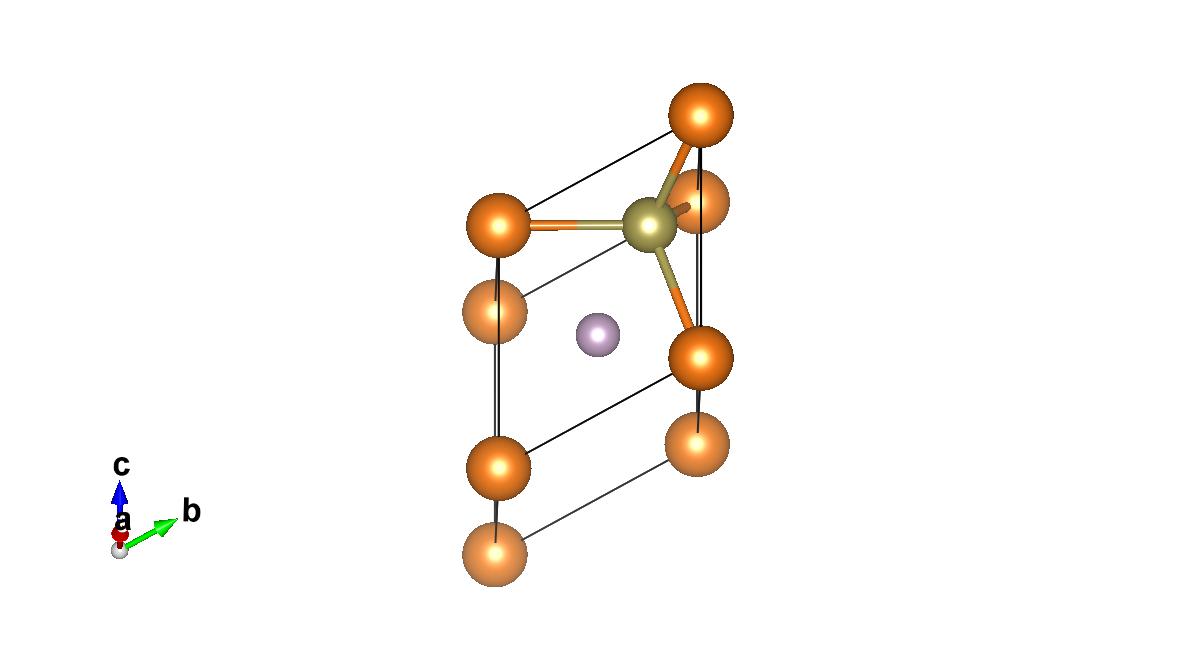}
        \caption{agm004238592}
        \label{fig:agm004238592}
    \end{subfigure}\hfill%
    % Second figure
    \begin{subfigure}[b]{0.3\textwidth}
        \includegraphics[width=\textwidth]{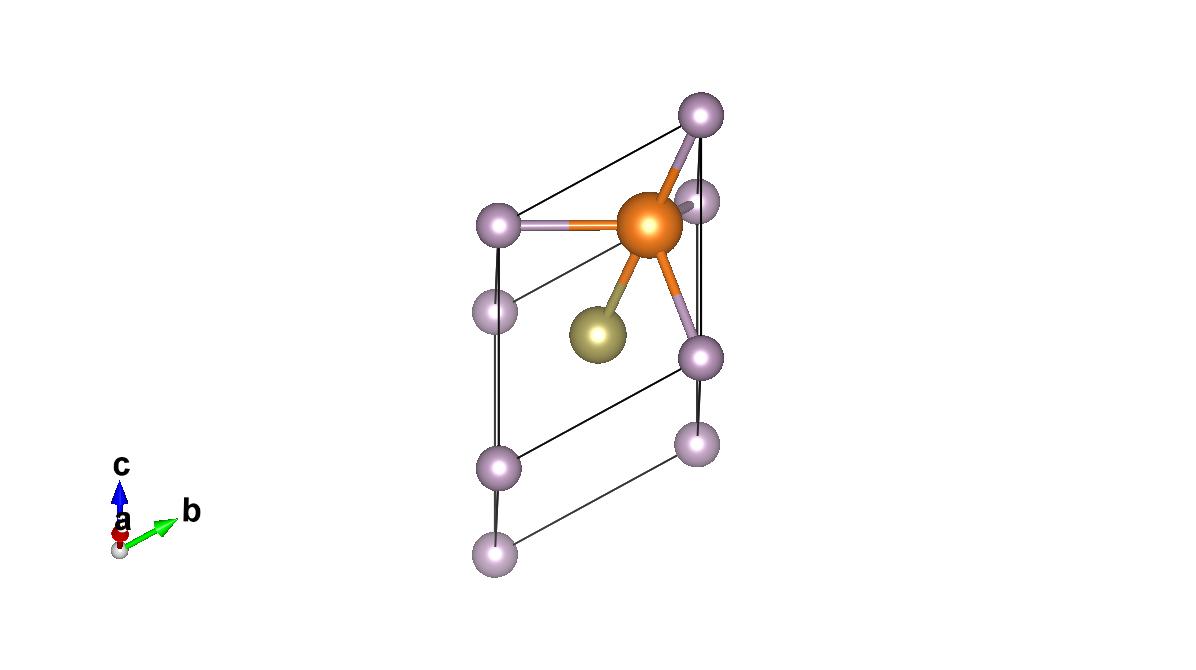}
        \caption{agm004238612}
        \label{fig:agm004238612}
    \end{subfigure}\hfill%
    % Third figure
    \begin{subfigure}[b]{0.3\textwidth}
        \includegraphics[width=\textwidth]{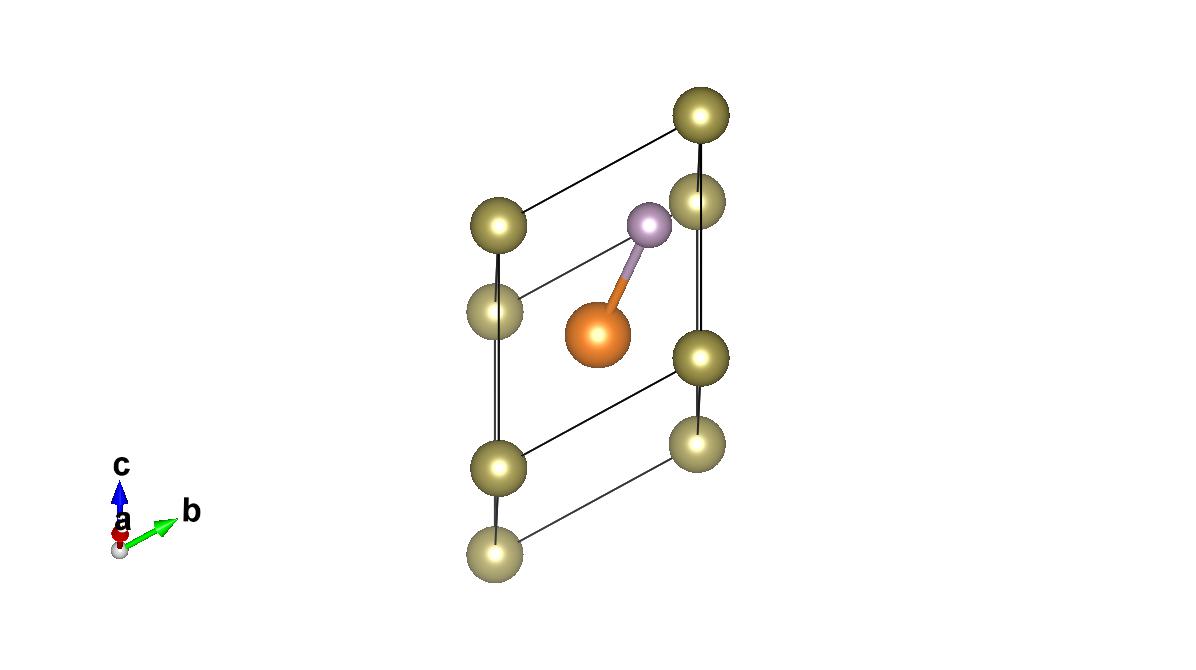}
        \caption{agm004238613}
        \label{fig:agm004238613}
    \end{subfigure}
    \caption{Three Alexandria structures with same Alexandria ID.}
    \label{fig:entalpic_fingerprint_agm}
\end{figure}

An identified triple pair of duplicates within Alexandria include \textit{agm004238592}, \textit{agm004238613}, and \textit{agm004238612}, all materials with a composition of \ce{MgTeP}. All three have rhombohedral lattices but with slightly different lattice parameters, with side length ranging from 4.8, to 4.71 to 4.70\AA{}\ref{fig:entalpic_fingerprint_agm}. The last two thus have similar densities while the first has a decreased density. In fact, all three are simply site permutations, with one have the Mg site at the origin, another having the Te site at the origin and the final having the P site at the origin. All three, as inputted in the database have norm of force vectors of 0$ev/A$ for all sites. Their energies differ slightly, from -3.03, to -3.19, to -3.31$eV/\AA{}$. All 3 are detected to be the same by Pymatgen's Structure Matcher. We can safely assume that these are simply permutations of the same materials with different levels of relaxation. For any kind of thermodynamic analysis, likely only the lowest energy one, with P centered at the origin, and the highest density, and smallest lattice vector is important. Such a permutation in Alexandria could have occured due to the prototype-library generation method, whereby considering a prototype ABC, and considering all potential element combination either two prototypes converged to the same structure or permutations was not accounted for in generating prototypes.

\clearpage
\subsection{Case study: specific hashing example for pair of duplicate found between Alexandria and Materials Project}

\begin{figure}[htbp]
    \centering
    \begin{subfigure}[b]{0.45\textwidth}
        \includegraphics[width=\textwidth]{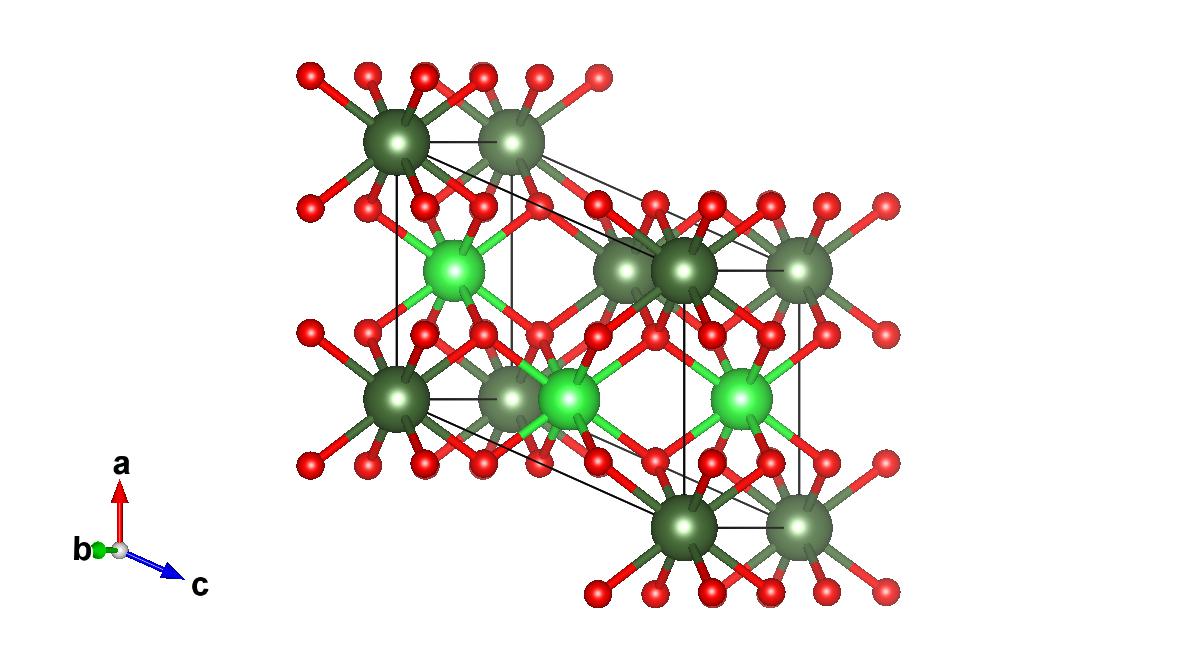}
        \caption{As inputted, Alexandria structure (agm003740092)}
        \label{fig:agm003740092_db}
    \end{subfigure}
    \hfill
    \begin{subfigure}[b]{0.45\textwidth}
        \includegraphics[width=\textwidth]{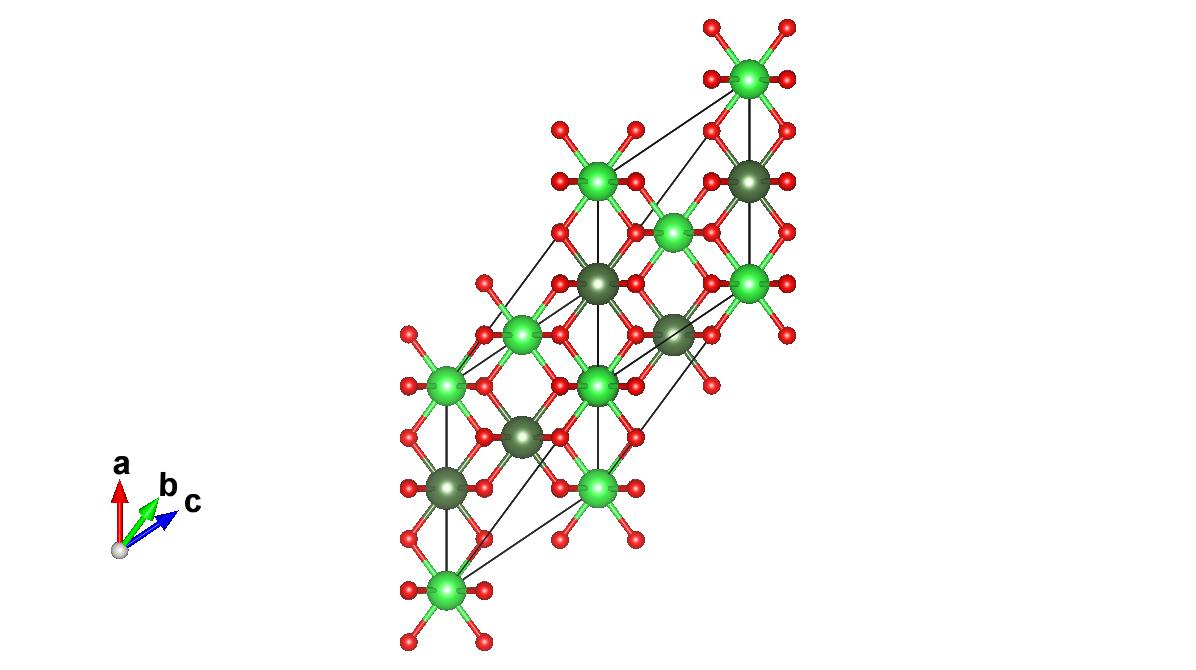}
        \caption{As inputted, Materials Project structure (mp-675871)}
        \label{fig:mp675871_db}
    \end{subfigure}

    \vskip\baselineskip

    \begin{subfigure}[b]{0.45\textwidth}
        \includegraphics[width=\textwidth]{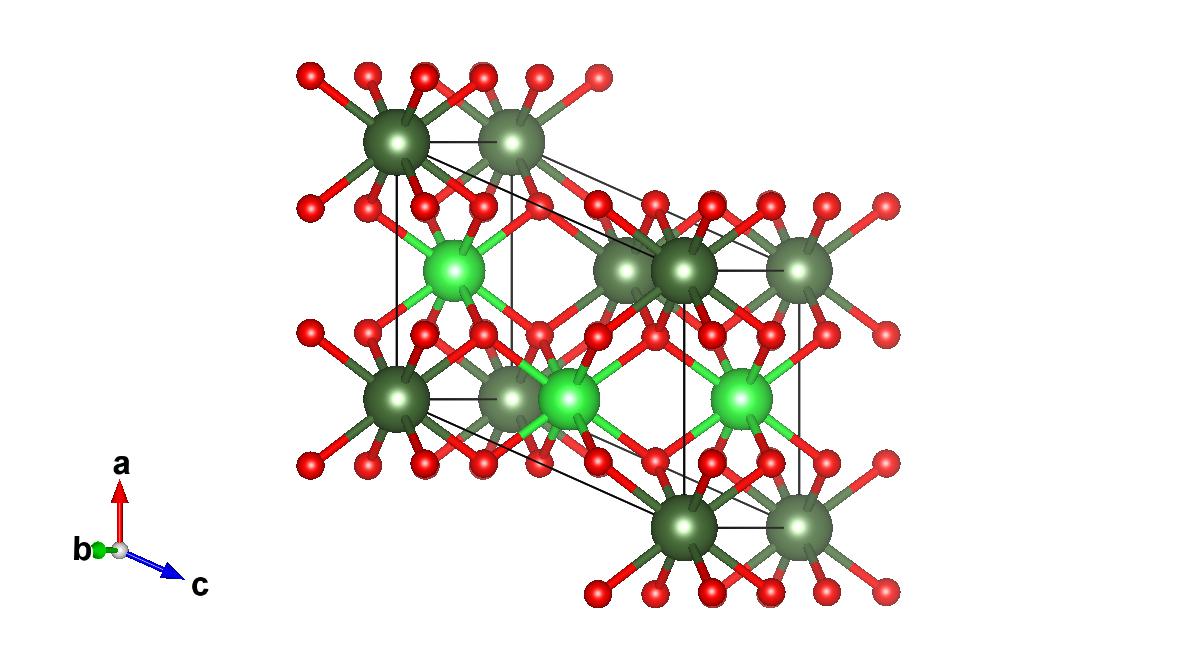}
        \caption{Niggli reduced Alexandria structure (agm003740092)}
        \label{fig:agm003740092_niggli}
    \end{subfigure}
    \hfill
    \begin{subfigure}[b]{0.45\textwidth}
        \includegraphics[width=\textwidth]{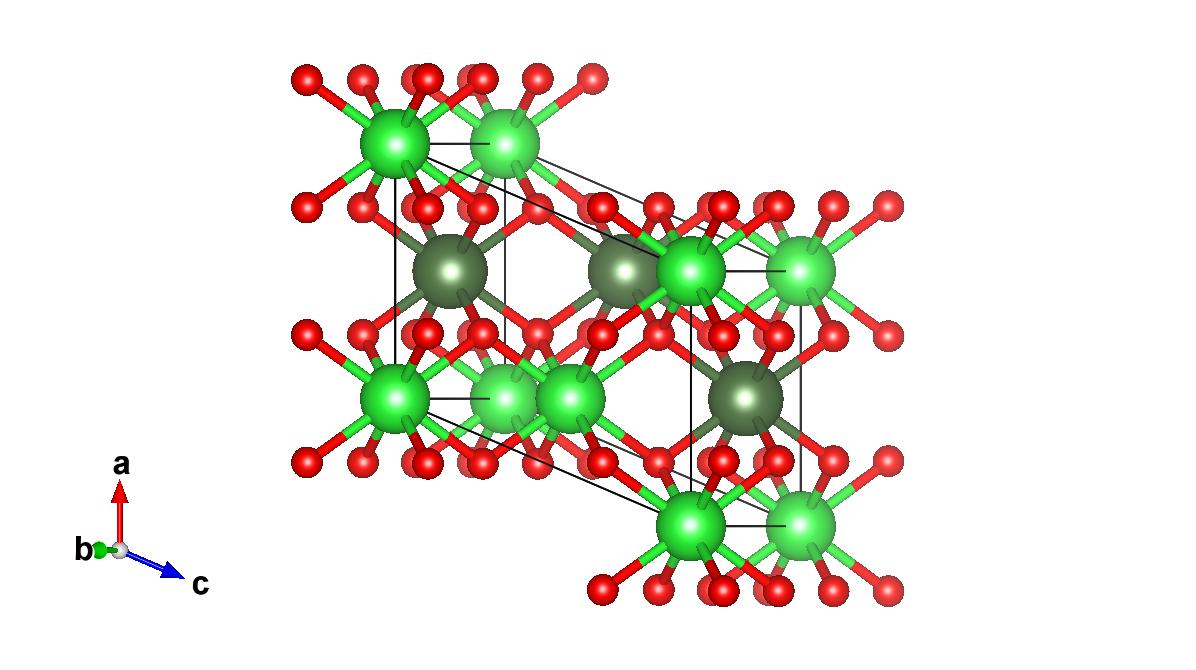}
        \caption{Niggli reduced Materials Project structure (mp-675871)}
        \label{fig:mp675871_niggli}
    \end{subfigure}
    
    \caption{As inputted and Niggli reduction of a selected material in LeMat-Bulk having the same fingerprint between Alexandria and Materials Project.}
    \label{fig:fourfigures}
\end{figure}

An identified pair of material which had a large energy difference between Alexandria and Materials Project was Material ID (\textit{agm003740092}) and Material ID (\textit{mp-675871}), materials with reduced composition \ce{Er2Pa2O8}, both calculated via PBE in each databases. In each database the material is presented different. Both materials are Spacegroup 141 by SPGLib, however their origin has been shifted: in Alexandria the material's inversion center is located at [0.0,  0.5,  0.25] (Fig. \ref{fig:agm003740092_db}) while in Materials Project the inversion center is at the origin ([0.0, 0.0, 0.0]) (Fig. \ref{fig:mp675871_db}). Additionally the sites are permuted, between Alexandria and Materials Project, the a sites and b sites have been permuted. The lattices are also represented differently, Materials Project lists the structure in the primitive configuration, while Alexandria has the Niggli reduction of the structure. When converting both structures to their Niggli reduction (Fig. \ref{fig:agm003740092_niggli} and Fig. \ref{fig:mp675871_niggli}), the lattice is additionally stretched. Materials Project has a density of 13.86$g/cm^3$ while Alexandria has a density of 9.93$g/cm^3$. Energy wise they have a difference of energy/per atom of over 5eV. When comparing the max of the norm of the forces on each sites between both structures is high for Materials Project (0.63$eV/\AA{}^2$ and low for Alexandria, $0.001eV/\AA{}^2$. For both cases, Pymatgen's StructureMatcher detects these two structures as inputted to be the same structure. These two structures underwent a DFT relaxation using the parameters previously described. After relaxation, the energy per atom of both structures approached -10eV/atom, and their densities relaxed to 10.4$g/cm^3$. Both energy and density values were closer to the Alexandria reported values (9.93$g/cm^3$ and 9.38$eV/atom$), though slightly different.

\section{Sample DFT input calculation parameters}
\begin{lstlisting}
ALGO = Normal
EDIFF = 2e-07
EDIFFG = 5e-07
ENAUG = 1360
ENCUT = 600
GGA = Ps
IBRION = 2
ISIF = 3
ISMEAR = 0
ISPIN = 2
ISYM = -1
IVDW = 11
LAECHG = True
LASPH = True
LCHARG = False
LELF = False
LMAXMIX = 6
LMIXTAU = True
LORBIT = 11
LREAL = False
LVTOT = True
LWAVE = False
MAGMOM = 8*0.6
NELM = 200
NSW = 400
POTIM = 0.06
PREC = Accurate
SIGMA = 0.01
\end{lstlisting}

\clearpage
\section{Biases in Databases}
All quantum chemistry database will have be exposed to chemical bias.
\begin{figure}
\centering
\includegraphics[height=10cm]{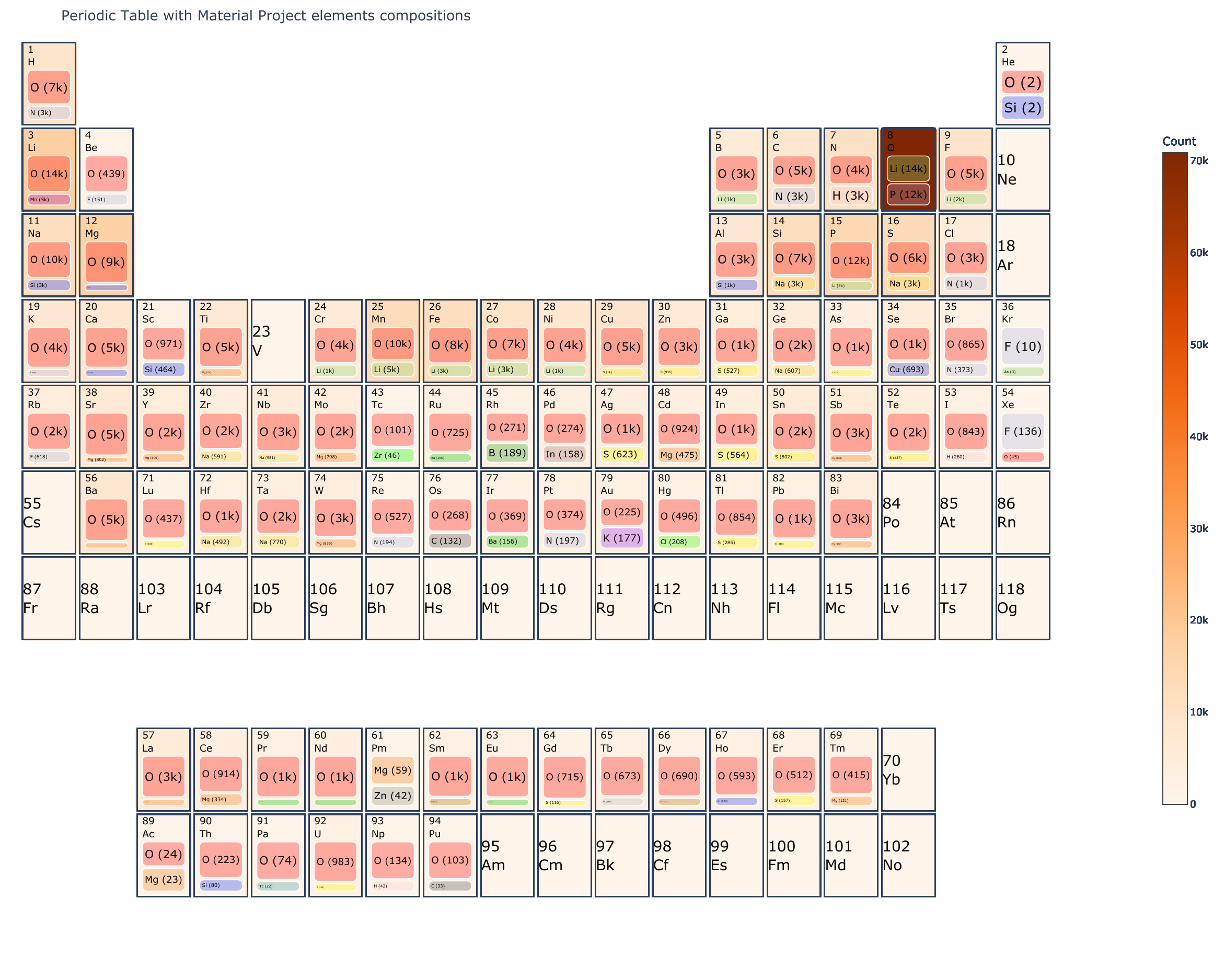}%
\caption{2D block chart demonstrating chemical bias in Materials Project. Outside blocks of the chart are colored based on number of materials which contain this element. Inside block area is based on relative top two other elements within the composition space of the outside block. A larger block area has more relative materials. Colors for inner block are set by element rather than by number of materials containing such elements. Materials Project shows increased entries for battery and oxide materials. \label{fig:bias-mp}}
\end{figure}

\begin{figure}
\centering
\includegraphics[height=10cm]{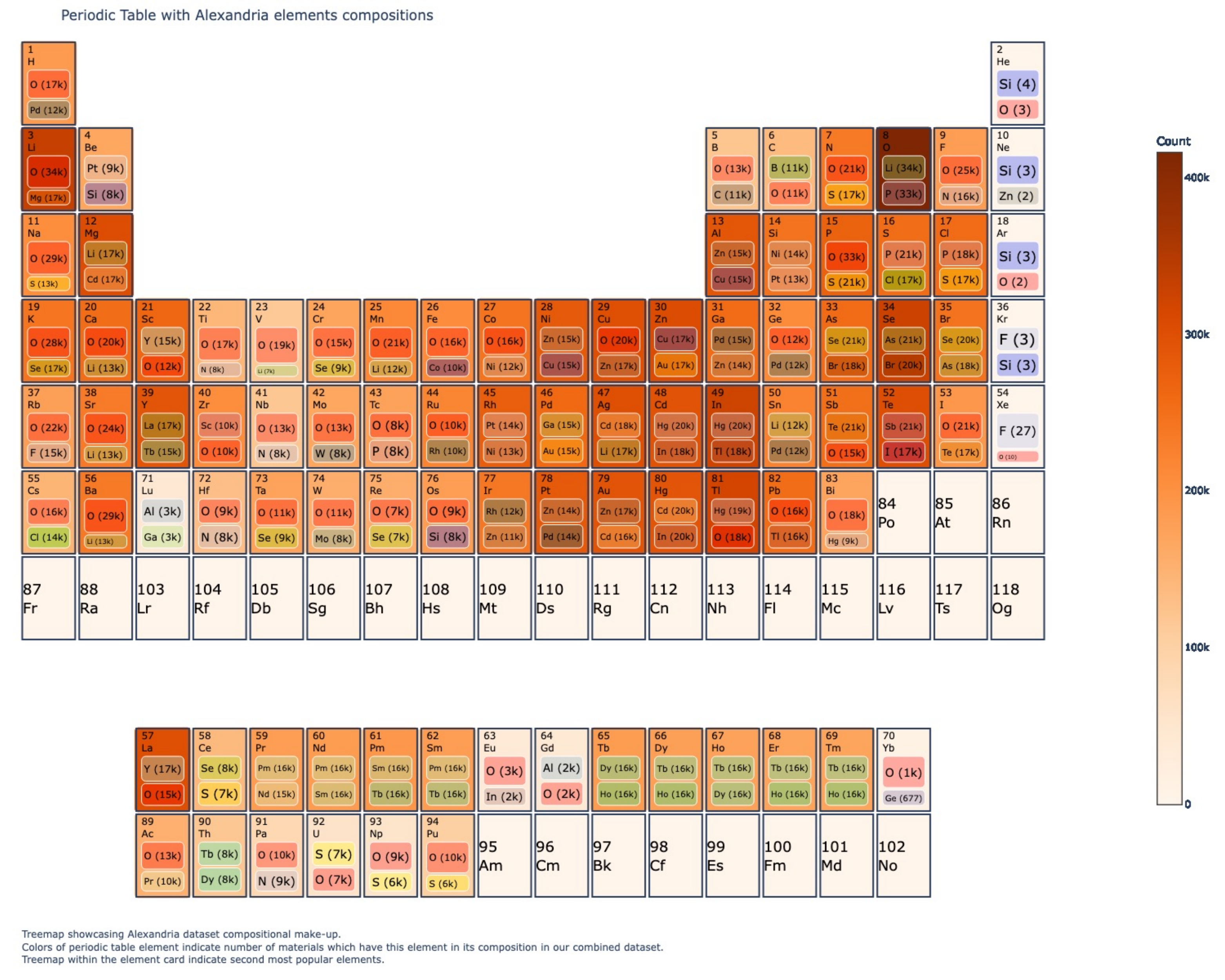}%
\caption{Alexandria shows increased entries for bimetallics\label{fig:bias-alexandria}}
\end{figure}

\begin{figure}
\centering
\includegraphics[height=10cm]{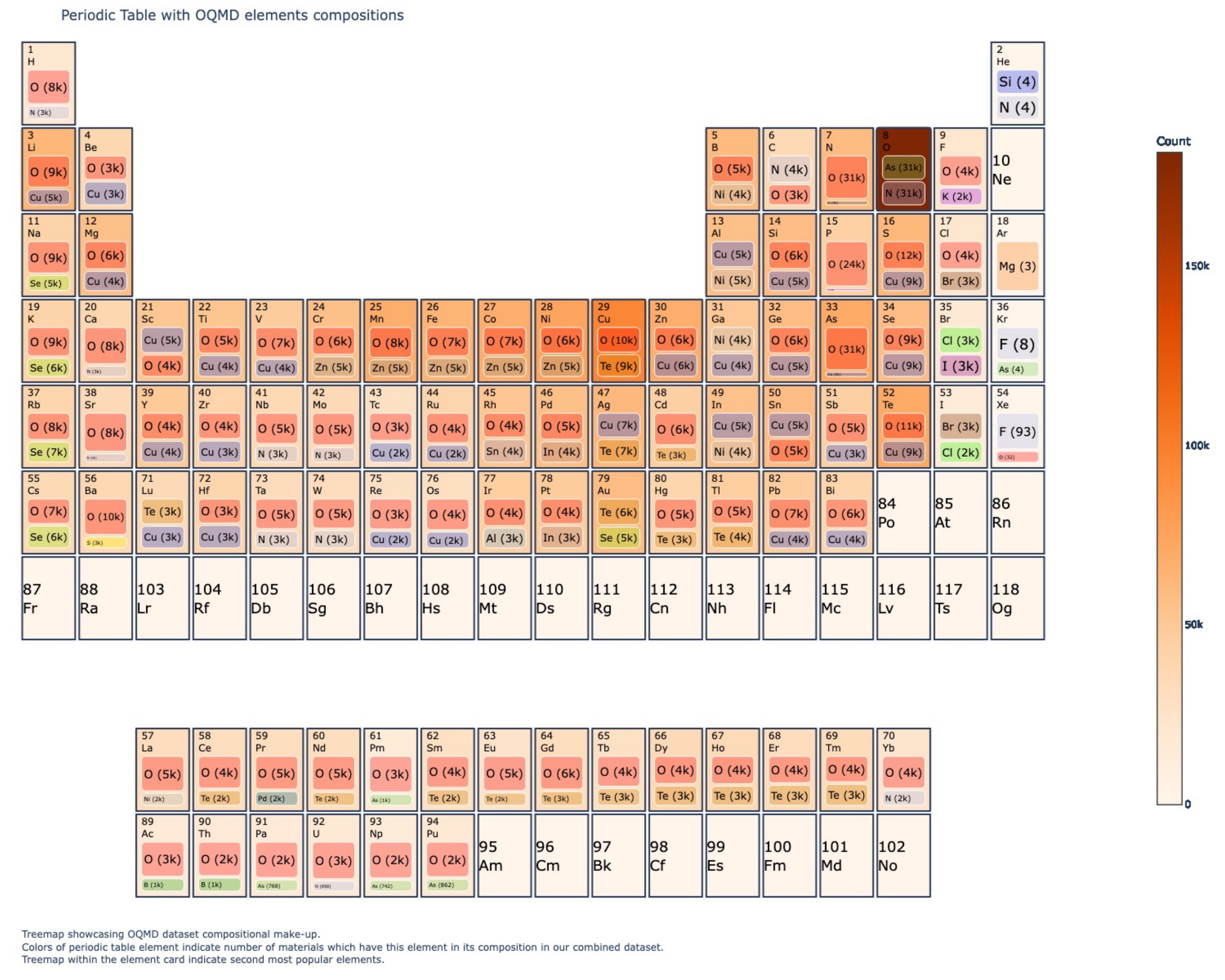}%
\caption{OQMD chemical composition distribution.\label{fig:bias-oqmd}}
\end{figure}

\clearpage
\section{Force-unconverged calculations}
\begin{figure}[!htb]
\centering
\includegraphics[height=6cm]{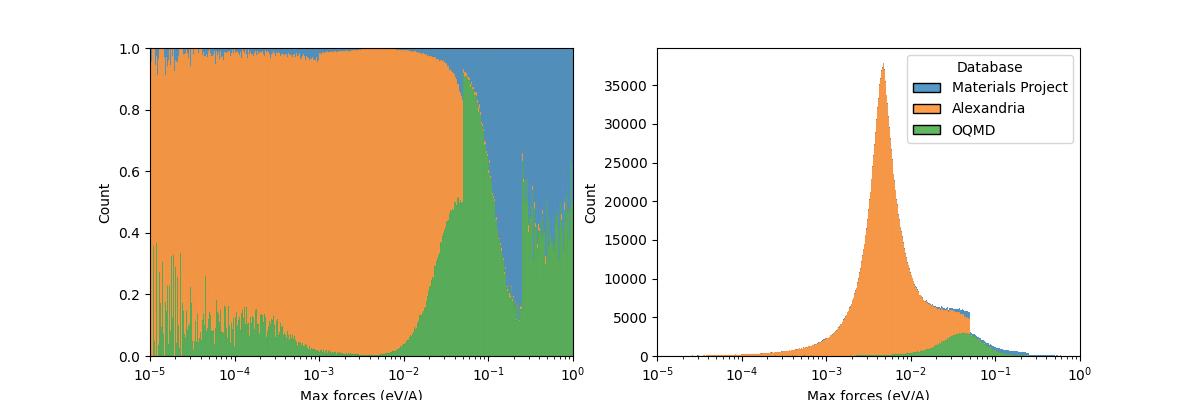}%
\caption{Comparing across databases for the max norm of the force vectors. Analysis shows most well converged structures by force appear to be those of Alexandria. Meanwhile a significant amount of structures with high force vectors (likely not fully converged structures) come from OQMD and Materials Project.}
\label{fig:max_force}
\end{figure}

\begin{figure}[!htb]
\centering
\includegraphics[height=6cm]{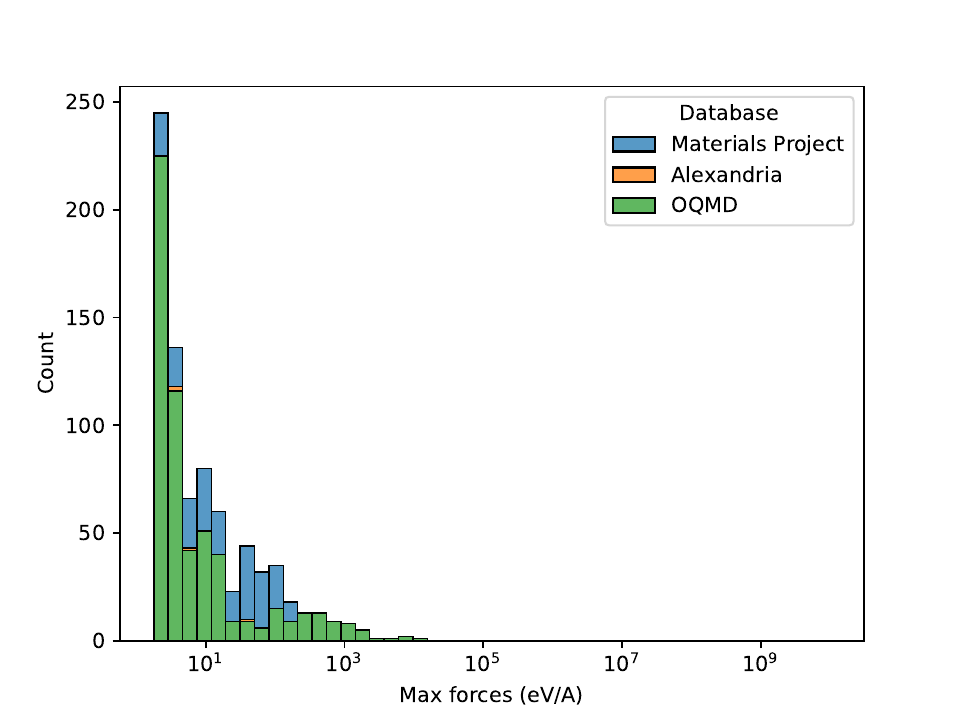}%
\caption{A significant amount of structures from both OQMD and Materials Project have very large force vectors.}
\label{fig:large_force}
\end{figure}

\clearpage
\section{Comparison of energy difference across same fingerprint materials across compatible and non-compatible database entries}

\begin{figure}
\centering
\includegraphics[width=0.6 \linewidth]{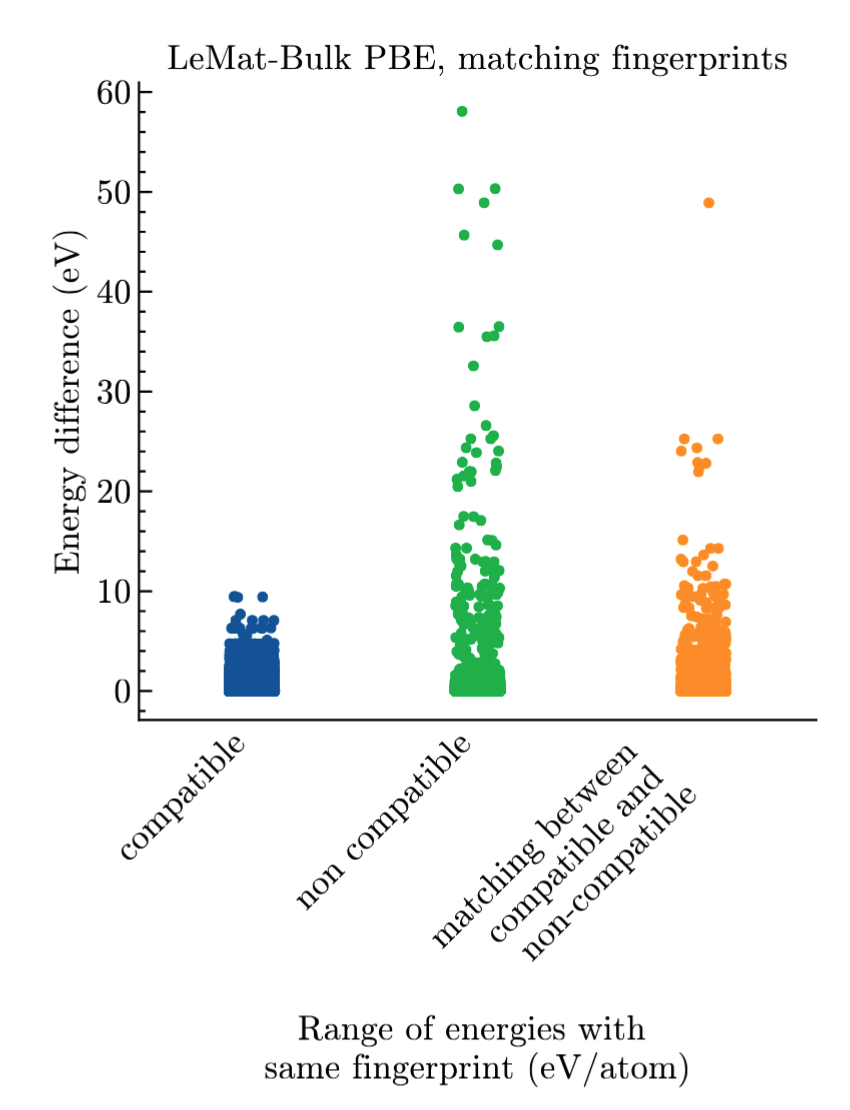}
\label{fig:energy_ranges}
\caption{Comparing energy range between compatible and non compatible entries for materials with the same fingerprints}
\end{figure}

\clearpage
\section{Most dissimilar structures with same calculated hashes}
\subsection{Most dissimilar by energy}
\begin{figure}[!htb]
\centering
\includegraphics{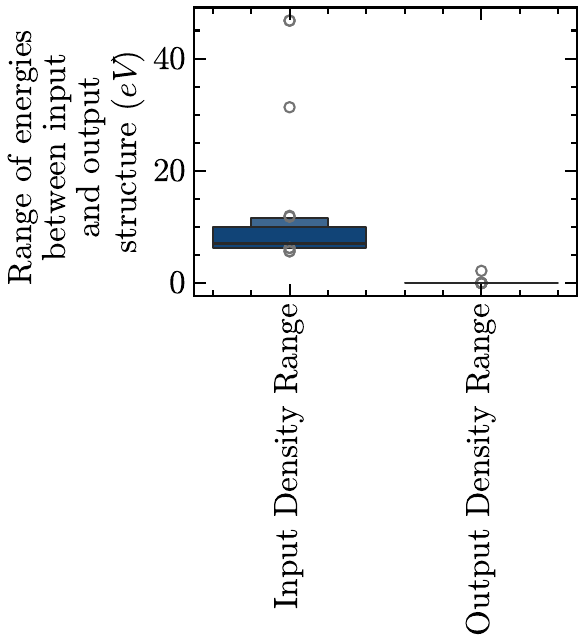}%
\caption{Comparing the input and output energies of most dissimilar structures by energy value in database with the same hash before and after relaxation.}
\label{fig:input_output_energy_most_dissimilar_energies}
\end{figure}

\begin{figure}[!htb]
\centering
\includegraphics{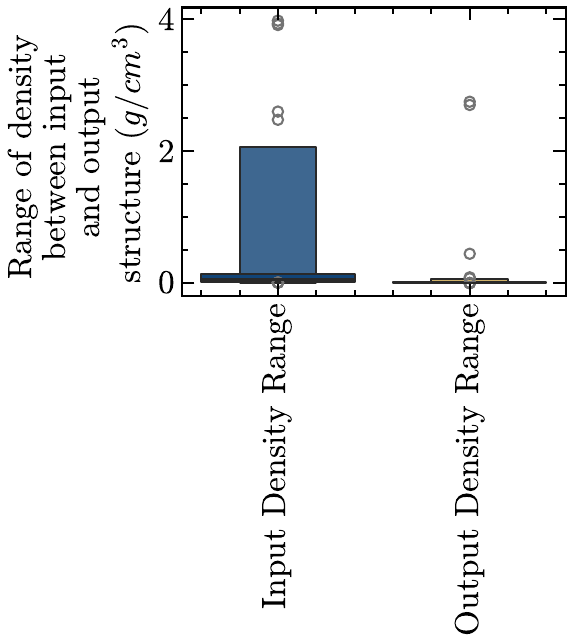}%
\caption{Comparing the input and output densities of most dissimilar structures by energy value in database with the same hash before and after relaxation.}
\label{fig:input_output_density_most_dissimilar_energies}
\end{figure}

When comparing the most dissimilar structures by energy value in database with the same hash before and after relaxation. After relaxation the energy range condenses to a mean of 7e-2 eV, with a minimum and maximum value of 8e-8 and 2.18eV. Prior to relaxation, the energy difference between those structures has a mean of 10eV with a minimum of 5.65eV and maximum of 46eV (Fig. \ref{fig:input_output_energy_most_dissimilar_energies}). Similarly after relaxation the density range condenses to a mean of 0.17 $g/cm^3$, with a minimum and maximum value of 7e-7 and 2.75$g/cm^3$ respectively. Prior to relaxation, the density difference between these structures had a mean of 0.55$g/cm^3$ with a minimum of 7.52e-2$g/cm^3$ and maximum of 3.98$g/cm^3$ 
(Fig. \ref{fig:input_output_density_most_dissimilar_energies}).

\clearpage
\subsection{Most dissimilar by GNN embeddings}
\begin{figure}[!htb]
\centering
\includegraphics{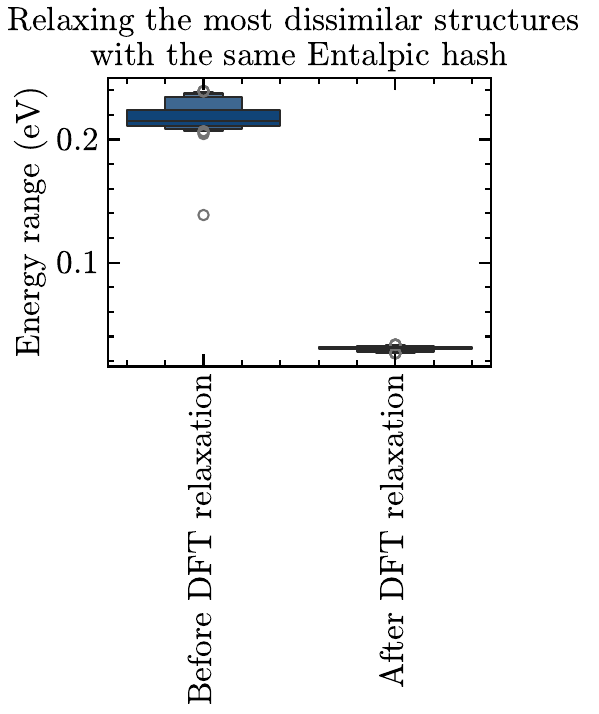}%
\label{fig:input_output_energy_most_dissimilar_gnn}
\caption{Comparing the input and output energies of most dissimilar structures by GNN embeddings in database with the same hash before and after relaxation.}
\end{figure}

\begin{figure}[!htb]
\centering
\includegraphics{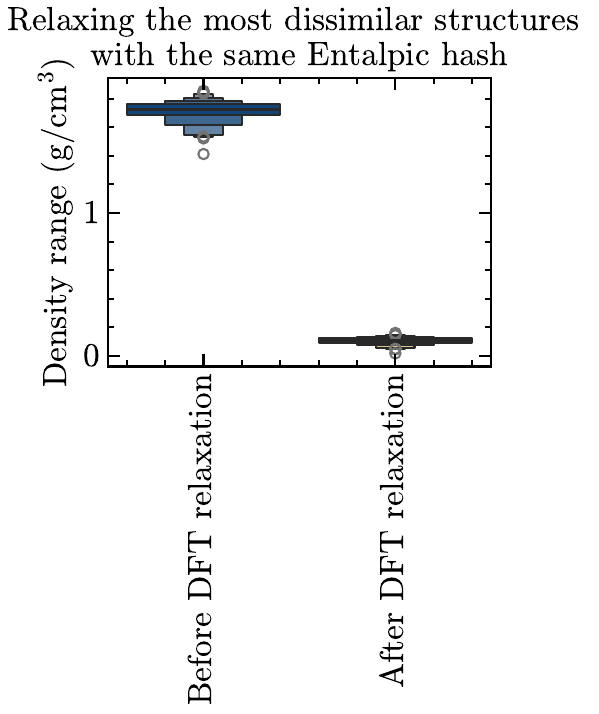}%
\label{fig:input_output_density_most_dissimilar_gnn}
\caption{Comparing the input and output densities of most dissimilar structures by GNN embeddings in database with the same hash before and after relaxation.}
\end{figure}

When comparing the most dissimilar structures by GNN embeddings in the database with the same hash before and after relaxation. After relaxation the energy range condenses to a mean of 0.030eV, with a minimum and maximum value of 0.026 and 0.033eV. Prior to relaxation, the energy difference between those structures has a mean of 0.21eV with a minimum of 0.13eV and maximum of 0.24eV (Fig. \ref{fig:input_output_energy_most_dissimilar_energies}). Similarly after relaxation the density range condenses to a mean of 0.10 $g/cm^3$, with a minimum and maximum value of 0.02 and 0.16$g/cm^3$ respectively. Prior to relaxation, the density difference between these structures had a mean of 1.71$g/cm^3$ with a minimum of 1.41$g/cm^3$ and maximum of 1.85$g/cm^3$ respectively 
(Fig. \ref{fig:input_output_density_most_dissimilar_energies}).

\clearpage
\section{Noisy structure vs. DFT Relaxation}
\begin{figure}[!htb]
\centering
\includegraphics{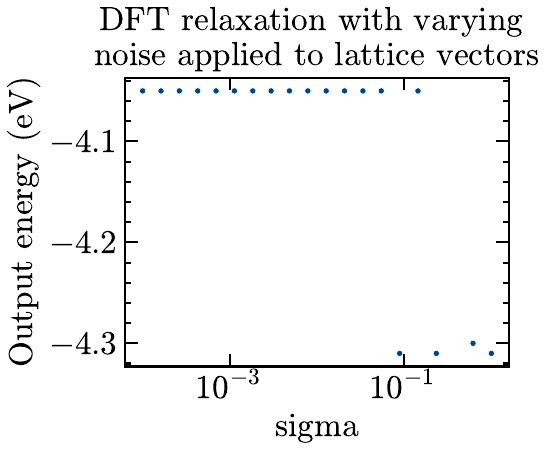}%
\label{fig:input_output_energy_most_dissimilar_gnn}
\caption{Noise to lattice vectors show a clear DFT sensitivity (relative to the DFT parameters used in this experiment) of around 0.1}
\end{figure}
\begin{figure}[!htb]
\centering
\includegraphics{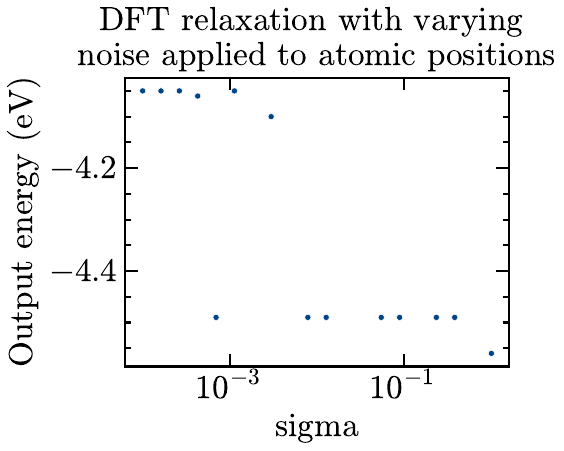}%
\label{fig:input_output_energy_most_dissimilar_gnn}
\caption{Noise to lattice vectors show a clear DFT sensitivity (relative to the DFT parameters used in this experiment) of around 0.1}
\end{figure}

Noise to lattice vectors show a clear DFT sensitivity (relative to the DFT parameters used in this experiment) of around 0.1. Meanwhile, noise to atomic positions are more sensitive, beginning to show clear relaxation differences in DFT around 0.01.

\clearpage
\section{Pairwise structures comparison, time of algorithms}

\begin{figure}[!htb]
\centering
\includegraphics{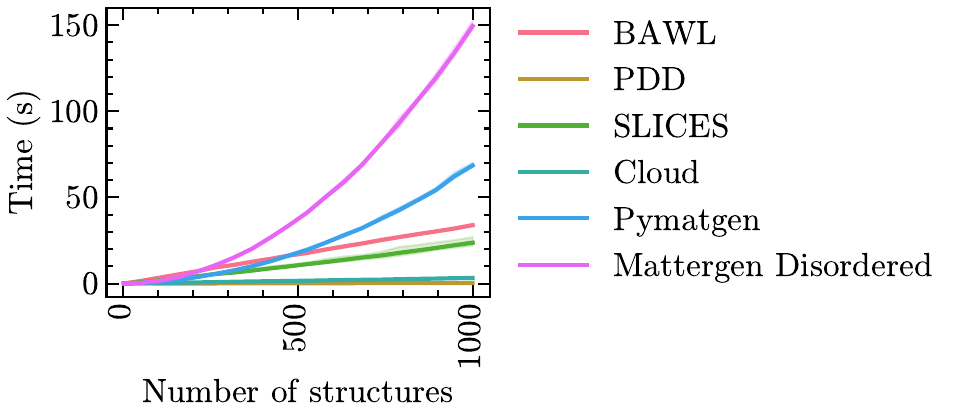}%
\caption{Time to pairwise compare set of structures}
\label{fig:pairwise_time}
\end{figure}

\begin{figure}[!htb]
\centering
\includegraphics{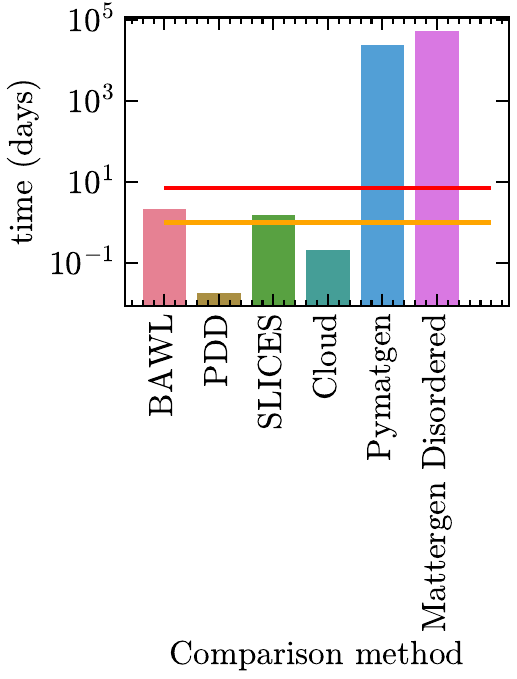}%
\caption{Extrapolated time to pairwise compare all materials in LeMat-Bulk}
\label{fig:pairwise_extrapolate_time}
\end{figure}

\clearpage
\section{Overall parameters used in LeMat-Bulk}

\begin{longtable}{llccc}
\caption{Pseudopotentials adopted in LeMat-Bulk}\label{tab:pseudopotentials}\\
\toprule
Element & LeMat-Bulk & Alexandria & Materials Project & OQMD \\
\midrule
\endfirsthead
\multicolumn{5}{c}{\tablename\ \thetable\ -- \textit{Continued from previous page}} \\
\toprule
Elements & LeMat-Bulk & Alexandria & Materials Project & OQMD \\
\midrule
\endhead
\midrule \multicolumn{5}{r}{\textit{Continued on next page}} \\
\endfoot
\bottomrule
\endlastfoot
Ac      & PAW\_PBE Ac 06Sep2000    & \checkmark & \checkmark & $\checkmark$ \\
Ag      & PAW\_PBE Ag 06Sep2000    & \checkmark & \checkmark & $\checkmark$ \\
Al      & PAW\_PBE Al 04Jan2001    & \checkmark & \checkmark & $\checkmark$ \\
Ar      & PAW\_PBE Ar 07Sep2000    & \checkmark & \checkmark & $\checkmark$ \\
As      & PAW\_PBE As 06Sep2000    & \checkmark & \checkmark & $\checkmark$ \\
At      & PAW\_PBE At\_d     & $\times$ \text{ (not in DB)} & $\times$ \text{ (not in DB)} & $\checkmark$ \\
Au      & PAW\_PBE Au 06Sep2000    & \checkmark & \checkmark & $\checkmark$ \\
Ba      & PAW\_PBE Ba\_sv 06Sep2000 & \checkmark & \checkmark & $\checkmark$ \\
Be      & PAW\_PBE Be\_sv 06Sep2000 & \checkmark & \checkmark & $\checkmark$ \\
Bi      & PAW\_PBE Bi 08Apr2002    & \checkmark & \checkmark & $\checkmark$ \\
B       & PAW\_PBE B 06Sep2000     & \checkmark & \checkmark & $\checkmark$ \\
Br      & PAW\_PBE Br 06Sep2000    & \checkmark & \checkmark & $\checkmark$ \\
Ca      & PAW\_PBE Ca\_sv 06Sep2000 & \checkmark & \checkmark & $\times$\text{ (Ca\_pv)} \\
Cd      & PAW\_PBE Cd 06Sep2000    & \checkmark & \checkmark & $\checkmark$ \\
Ce      & PAW\_PBE Ce 28Sep2000    & \checkmark & \checkmark & $\times$ \text{ (Ce\_3)}\\
Cl      & PAW\_PBE Cl 17Jan2003    & \checkmark & \checkmark & $\checkmark$ \\
Co      & PAW\_PBE Co 06Sep2000    & \checkmark & \checkmark & $\checkmark$ \\
C       & PAW\_PBE C 08Apr2002     & \checkmark & \checkmark & $\checkmark$ \\
Cr      & PAW\_PBE Cr\_pv 07Sep2000 & \checkmark & \checkmark & $\times$ \text{ (Ce\_3)} \\
Cs      & PAW\_PBE Cs\_sv 08Apr2002 & \checkmark & \checkmark & $\checkmark$ \\
Cu      & PAW\_PBE Cu\_pv 06Sep2000 & \checkmark & \checkmark & $\checkmark$ \\
Dy      & PAW\_PBE Dy\_3 06Sep2000  & \checkmark & \checkmark & $\checkmark$ \\
Er      & PAW\_PBE Er\_3 06Sep2000  & \checkmark & \checkmark & $\checkmark$ \\
Eu      & PAW\_PBE Eu 08Apr2002    & \checkmark & \checkmark & $\times$\text{ (Eu\_2)} \\
Fe      & PAW\_PBE Fe\_pv 06Sep2000 & \checkmark & \checkmark & $\checkmark$ \\
F       & PAW\_PBE F 08Apr2002     & \checkmark & \checkmark & $\checkmark$ \\
Ga      & PAW\_PBE Ga\_d 06Sep2000  & \checkmark & \checkmark & $\checkmark$ \\
Gd      & PAW\_PBE Gd 08Apr2002    & \checkmark & \checkmark & $\checkmark$ \\
Ge      & PAW\_PBE Ge\_d 06Sep2000  & \checkmark & \checkmark & $\checkmark$ \\
He      & PAW\_PBE He 05Jan2001    & \checkmark & \checkmark & $\checkmark$ \\
Hf      & PAW\_PBE Hf\_pv 06Sep2000 & \checkmark & \checkmark & $\checkmark$ \\
Hg      & PAW\_PBE Hg 06Sep2000    & \checkmark & \checkmark & $\checkmark$ \\
Ho      & PAW\_PBE Ho\_3 06Sep2000  & \checkmark & \checkmark & $\checkmark$ \\
H       & PAW\_PBE H 15Jun2001     & \checkmark & \checkmark & $\checkmark$ \\
In      & PAW\_PBE In\_d 06Sep2000  & \checkmark & \checkmark & $\checkmark$ \\
I       & PAW\_PBE I 08Apr2002     & \checkmark & \checkmark & $\checkmark$ \\
Ir      & PAW\_PBE Ir 06Sep2000    & \checkmark & \checkmark & $\checkmark$ \\
K       & PAW\_PBE K\_sv 06Sep2000  & \checkmark & \checkmark & $\checkmark$ \\
Kr      & PAW\_PBE Kr 07Sep2000    & \checkmark & \checkmark & $\checkmark$ \\
La      & PAW\_PBE La 06Sep2000    & \checkmark & \checkmark & $\checkmark$ \\
Li      & PAW\_PBE Li\_sv 23Jan2001 & \checkmark & \checkmark & $\checkmark$ \\
Lu      & PAW\_PBE Lu\_3 06Sep2000  & \checkmark & \checkmark & $\checkmark$ \\
Mg      & PAW\_PBE Mg\_pv 06Sep2000 & \checkmark & \checkmark & $\checkmark$ \\
Mn      & PAW\_PBE Mn\_pv 07Sep2000 & \checkmark & \checkmark & $\times$ \text{ (Mn)} \\
Mo      & PAW\_PBE Mo\_pv 08Apr2002 & \checkmark & \checkmark & $\checkmark$ \\
Na      & PAW\_PBE Na\_pv 05Jan2001 & \checkmark & \checkmark & $\checkmark$ \\
Nb      & PAW\_PBE Nb\_pv 08Apr2002 & \checkmark & \checkmark & $\checkmark$ \\
Nd      & PAW\_PBE Nd\_3 06Sep2000  & \checkmark & \checkmark & $\checkmark$ \\
Ne      & PAW\_PBE Ne 05Jan2001    & \checkmark & \checkmark & $\checkmark$ \\
Ni      & PAW\_PBE Ni\_pv 06Sep2000 & \checkmark & \checkmark & $\checkmark$ \\
N       & PAW\_PBE N 08Apr2002     & \checkmark & \checkmark & $\checkmark$ \\
Np      & PAW\_PBE Np 06Sep2000    & \checkmark & \checkmark & $\checkmark$ \\
O       & PAW\_PBE O 08Apr2002     & \checkmark & \checkmark & $\checkmark$ \\
Os      & PAW\_PBE Os\_pv 20Jan2003 & \checkmark & \checkmark & $\checkmark$ \\
Pa      & PAW\_PBE Pa 07Sep2000    & \checkmark & \checkmark & $\checkmark$ \\
Pb      & PAW\_PBE Pb\_d 06Sep2000  & \checkmark & \checkmark & $\checkmark$ \\
Pd      & PAW\_PBE Pd 05Jan2001    & \checkmark & \checkmark & $\checkmark$ \\
Pm      & PAW\_PBE Pm\_3 07Sep2000  & \checkmark & \checkmark & $\checkmark$ \\
P       & PAW\_PBE P 17Jan2003     & \checkmark & \checkmark & $\checkmark$ \\
Po       & PAW\_PBE Po     & $\times$ \text{ (not in DB)} & $\times$ \text{ (not in DB)} & $\checkmark$ \\
Pr      & PAW\_PBE Pr\_3 07Sep2000  & \checkmark & \checkmark & $\checkmark$ \\
Pt      & PAW\_PBE Pt 05Jan2001    & \checkmark & \checkmark & $\checkmark$ \\
Pu      & PAW\_PBE Pu 06Sep2000    & \checkmark & \checkmark & $\checkmark$ \\
Rb      & PAW\_PBE Rb\_sv 06Sep2000 & \checkmark & \checkmark & $\checkmark$ \\
Re      & PAW\_PBE Re\_pv 06Sep2000 & \checkmark & \checkmark & $\checkmark$ \\
Rh      & PAW\_PBE Rh\_pv 06Sep2000 & \checkmark & \checkmark & $\times$ \text{ (Rh)} \\
Ru      & PAW\_PBE Ru\_pv 06Sep2000 & \checkmark & \checkmark & $\times$ \text{ (Ru)} \\
Sb      & PAW\_PBE Sb 06Sep2000    & \checkmark & \checkmark & $\checkmark$ \\
Sc      & PAW\_PBE Sc\_sv 07Sep2000 & \checkmark & \checkmark & $\checkmark$ \\
Se      & PAW\_PBE Se 06Sep2000    & \checkmark & \checkmark & $\checkmark$ \\
Si      & PAW\_PBE Si 05Jan2001    & \checkmark & \checkmark & $\checkmark$ \\
Sm      & PAW\_PBE Sm\_3 07Sep2000  & \checkmark & \checkmark & $\checkmark$ \\
Sn      & PAW\_PBE Sn\_d 06Sep2000  & \checkmark & \checkmark & $\checkmark$ \\
S       & PAW\_PBE S 17Jan2003     & \checkmark & \checkmark & $\checkmark$ \\
Sr      & PAW\_PBE Sr\_sv 07Sep2000 & \checkmark & \checkmark & $\checkmark$ \\
Ta      & PAW\_PBE Ta\_pv 07Sep2000 & \checkmark & \checkmark & $\checkmark$ \\
Tb      & PAW\_PBE Tb\_3 06Sep2000  & \checkmark & \checkmark & $\checkmark$ \\
Tc      & PAW\_PBE Tc\_pv 06Sep2000 & \checkmark & \checkmark & $\checkmark$ \\
Te      & PAW\_PBE Te 08Apr2002    & \checkmark & \checkmark & $\checkmark$ \\
Th      & PAW\_PBE Th 07Sep2000    & \checkmark & \checkmark & $\checkmark$ \\
Ti      & PAW\_PBE Ti\_pv 07Sep2000 & \checkmark & \checkmark & $\times $ \text{ (Ti)} \\
Tl      & PAW\_PBE Tl\_d 06Sep2000  & \checkmark & \checkmark & $\checkmark$ \\
Tm      & PAW\_PBE Tm\_3 20Jan2003  & \checkmark & \checkmark & $\checkmark$ \\
U       & PAW\_PBE U 06Sep2000     & \checkmark & \checkmark & $\checkmark$ \\
V       & PAW\_PBE V\_sv 07Sep2000  & \checkmark & $\times$ \text{ (V\_pv)} & $\times$ \text{(V)} \\
W       & PAW\_PBE W\_pv 06Sep2000  & \checkmark & \checkmark & $\checkmark$ \\
Xe      & PAW\_PBE Xe 07Sep2000    & \checkmark & \checkmark & $\checkmark$ \\
Yb      & PAW\_PBE Yb 24Feb2003    & \checkmark & $\times$ \text{ (Yb\_3)} & $\times $\text{ (Yb\_2)} \\
Y       & PAW\_PBE Y\_sv 06Sep2000  & \checkmark & \checkmark & $\checkmark$ \\
Zn      & PAW\_PBE Zn 06Sep2000    & \checkmark & \checkmark & $\checkmark$ \\
Zr      & PAW\_PBE Zr\_sv 07Sep2000  & \checkmark & \checkmark & $\checkmark$ \\
\end{longtable}

\clearpage
\begin{table}[]
\begin{tabular}{lll}
\toprule
\textbf{Element} & \textbf{System} & \textbf{Hubbard U (eV)} \\
\midrule

Co 
& Oxides, Fluorides 
& 3.32  \\

Cr 
& Oxides, Fluorides 
& 3.7  \\

Fe 
& Oxides, Fluorides 
& 5.3  \\

Mn 
& Oxides, Fluorides
& 3.9  \\

Mo 
& Oxides, Fluorides 
& 4.38  \\

Ni 
& Oxides, Fluorides
& 6.2  \\

V 
& Oxides, Fluorides 
& 3.25 \\

W 
& Oxides, Fluorides
& 6.2 \\

\bottomrule
\end{tabular}
\caption{Hubbard-U values adopted in LeMat-Bulk}\label{tab:hubbardu}
\label{tab:calibrated-u-values}
\end{table}

Materials Project and Alexandria relaxed materials with at least a 1,000 kpoint density, while OQMD used a much denser 4,000 points criteria.

\section{LeMat-Bulk Statistic}

\begin{table}[!htbp]
\caption{Comparison of Database Sizes}
\begin{tabular}{lrr}
\toprule
\textbf{Database} & \textbf{Number of materials} & \textbf{Number of structures*} \\
\midrule
Materials Project & 148,453 & 189,403 \\
Alexandria & 4,635,066 & 5,459,260 \\
OQMD & 1,076,926 & 1,076,926 \\
LeMaterial (All) & 5,860,446 & 6,725,590 \\
LeMaterial (Compatible, PBE) & 5,335,299 & 5,335,299 \\
LeMaterial (Compatible, PBESOL) & 447,824 & 447,824 \\
LeMaterial (Compatible, SCAN) & 422,840 & 422,840 \\
\bottomrule
\end{tabular}
\label{tab:database-sizes}
\end{table}

\section{Bader charge calculations for Materials Project}

For each calculation, we associated CHGCAR files with their corresponding AECCAR0 and AECCAR2 files based on their task identification numbers. The reference charge density was computed by summing the AECCAR0 and AECCAR2 files using the charge density summation script provided by the VASP Transition State Tools (VTST)~\citep{utexasTransitionState}. Bader charge calculations~\citep{tang2009grid} were performed on all files with matching AECCAR's and CHGCAR's, and the results were output as JSON files containing decorated structure data. Task IDs were cross-referenced with the identifiers in our existing database, ensuring that the structural coordinates matched and that the site orders were consistent.

\section{Multiple calculations per materials}
Alexandria is a materials database where each calculation corresponds to a single material with a unique material ID. In contrast, the Materials Project and OQMD function as calculation repositories, associating multiple calculations (e.g., structure relaxation, static calculations, band structure computations) with a single material ID. This distinction introduces additional complexity when integrating data from these repositories. For OQMD and Materials Project, a static calculation is typically performed following structure relaxation, utilizing a denser k-point grid~\citep{saal2013materials}. For \textit{LeMat-Bulk}, we selected the outputs of the static calculations to extract energy, forces, magnetic moments, charge information, and the final structure.

\section{Code availability}
The dynamic nature of these databases requires continuous updates. Between the initial release of the dataset described in this work and the preparation of this manuscript, Alexandria added approximately 125,000 materials, and the Materials Project introduced a new database release. Further, Materials Project has a practice of re-running calculations with updated methodologies and improved parameters. In light of this, we focus on providing a methodology and accompanying code for database integration rather than a static dataset.

The code for database integration can be found here: \url{https://github.com/LeMaterial/lematerial-fetcher}. The code for structure fingerprint and similarity benchmarks can be found here: \url{https://github.com/LeMaterial/material-hasher}

\section{Comparing previous methodologies for structure similarity comparison and structure fingerprint generation}
Several previous methods have addressed structure matching and de-duplication. For example, Pymatgen's StructureMatcher \citep{ong2015materials} defined novelty based on a combination of angle, distance and symmetry tolerance, and provides for a combinatorial approach to compare new structures against all existing entries. While effective for small datasets, this approach becomes computationally prohibitive as databases grow. In building the Alexandria database, Alexandria used a criteria based on identical space groups, compositions, and energy differences, but similarly, its efficiency decreases with increasing database size.

Clusterization techniques, such as those used in C2DB \citep{gjerding2021recent}, incorporate Pareto analysis to reduce redundancy but do not provide fast hash-like lookups. In molecular databases, hashing methods like SMILES \citep{weininger1988smiles} and InChI \citep{heller2013inchi} offer string representations for rapid structure retrieval. Efforts to extend similar approaches to crystals have resulted in methods like SLICES \citep{xiao2023invertible} and its modified version in the CLOUD paper \citep{xu2024cloud}. However, these methods lack extensive benchmarking.

Other notable contributions include PDD vectorization \citep{widdowson2022resolving} and graph neural networks (GNNs) \citep{liao2023equiformerv2}, which encode crystal structures into a vectorized representations. While these methods offer promising alternatives for structure matching, they typically rely on arbitrary cutoffs for similarity and lack standardized benchmarks for performance evaluation. Moreover, their effectiveness in handling disordered structures remains unexplored. Additionally, both suffer from quadratic scale for comparing structures due to their need to combinatorial approach to comparison. AFLOW's XtalFinder \citep{hicks2021aflow} introduces a prototype-matching strategy that computes similarity between input structures and known crystal prototypes, but its dependence on a predefined prototype library limits its utility for novel material discovery.

Another notable similarity contribution from Therien et. al~\citep{therrien2020matching} involves matching crystals atoms to atoms.

Among fingerprint techniques, SLICES has a distinct advantage: it provides an invertible string representation of a crystal structure, effectively serving as both a fingerprint and a compression algorithm. This capability, unique among the methods tested in this study, allows for the recreation of bulk crystal structures directly from the fingerprint string. 

Among other similarity methods include notable contributions from Organov et. al~\citep{oganov2009quantify} which modifies previous work where radial distribution functions along with ion-specific information are utilized to construct a structure fingerprint. However, as noted in the study, this fingerprint is not sensitive to atomic permutations. Zhu et. al~\citep{zhu2016fingerprint} created a fingerprint based on the eigenvalues of a matrix  from an atom centered Gaussian basis set. They compared structures using the Euclidean norm of their differences. Sandip et. al.~\citep{de2016comparing} proposed a SOAP~\citep{bartok2013representing}-based kernel for a structure, which can then be compared by taking the dot product (overlap) between two structures. Yang et. al.~\citep{yang2014proposed} makes use of a bonding algorithm (O'Keef) to construct a vector based on neighboring atoms along a central ions, and uses a sigmoid function to score similar structures. While important, these structure similarity methods are beyond the scope of this comparison study.

\end{document}